\documentclass[useAMS,usenatbib]{mn2e}
\usepackage{graphicx}  
\usepackage{float}     
\usepackage{amsmath}
\usepackage{longtable}
\usepackage{multicol}
\usepackage[T1]{fontenc}
\usepackage{aecompl}

\title[Grain motion effects on astrochemistry]
      {Effects of turbulent dust grain motion to interstellar chemistry}

\author[Ge, He \& Yan]
{ J.X. Ge$^{1,2}$,
  J.H. He$^{1}$,
  H.R. Yan$^{3,4}$  \\
$^{1}$ Key Laboratory for the Structure and Evolution of Celestial Objects,
  Yunnan observatories, Chinese Academy of Sciences, \\
  P.O. Box 110, Kunming, 650011, Yunnan Province, PR China. E-mail: jinhuahe@ynao.ac.cn\\
$^{2}$ Chinese Academy of Sciences University, PR China. E-mail: gejixing@ynao.ac.cn\\
$^{3}$ The Kavli Institute for Astronomy and Astrophysics at Peking University, Beijing, PR China.\\
$^{4}$ DESY \& University of Potsdam, Germany. E-mail: huirong.yan@desy.de\\
}

\begin{document}


\pagerange{\pageref{firstpage}--\pageref{lastpage}} \pubyear{2014}

\maketitle

\label{firstpage}

\begin{abstract}
Theoretical studies have revealed that dust grains are usually moving fast through the turbulent interstellar gas, which could have significant effects upon interstellar chemistry by modifying grain accretion. This effect is investigated in this work on the basis of numerical gas-grain chemical modeling. Major features of the grain motion effect in the typical environment of dark clouds (DC) can be summarised as follows: 1) decrease of gas-phase (both neutral and ionic) abundances and increase of surface abundances by up to 2-3 orders of magnitude; 2) shifts of the existing chemical jumps to earlier evolution ages for gas-phase species and to later ages for surface species by factors of about ten; 3) a few exceptional cases in which some species turn out to be insensitive to this effect and some other species can show opposite behaviors too. These effects usually begin to emerge from a typical DC model age of about $10^5$\,yr. The grain motion in a typical cold neutral medium (CNM) can help overcome the Coulomb repulsive barrier to enable effective accretion of cations onto positively charged grains. As a result, the grain motion greatly enhances the abundances of some gas-phase and surface species by factors up to 2-6 or more orders of magnitude in the CNM model. The grain motion effect in a typical molecular cloud (MC) is intermediate between that of the DC and CNM models, but with weaker strength. The grain motion is found to be important to consider in chemical simulations of typical interstellar medium.

\end{abstract}


\section{Introduction}
Astrochemistry is important in understanding of physical and chemical properties of the interstellar medium. The role of dust grains in the chemistry of cold interstellar gas was recognised since 1970s \citep[e.g.,][]{goul1963,holl1971} and the gas-grain chemistry began to develop \citep[e.g.,][]{tiel1982,hase1992}. The art of astrochemical modeling has been improved much in various aspects since then. For example, the surface reactions and chemical effects of grain accretion and desorption was explored in more details by \citet{schu1991,hase1993a,viti1999} and more recently by \citet{dalg2006,garr2006,vasy2013,hinc2015}, etc.; the chemical and physical nature of dust grains and its impact to astrochemistry was investigated by \citet{hase1993b,omon1986,lepp1988,wake2008,kalv2010,acha2011,lebo2012,garr2013a,garr2013b}, etc. The discrete nature of the grain chemistry problem was also addressed and new simulation approaches such as Monte Carlo method \citep[e.g.,][]{case1998,char1998,chan2005,chan2006,chan2007,vasy2009,chan2012,garr2013b,vasy2013}, master equation \citep{biha2001,stan2002}, moment equation \citep{barz2007} and modified rate equation \citep{char1997,case1998,garr2008,garr2009} have been proposed. One more such improvement for the chemical effect of turbulent grain motions will be presented in this work.

In a standard gas-grain chemical model, the accretion of gas onto grains is usually considered by assuming standstill grains in homogeneous gas so that the accretion is solely determined by the thermal motion of gas-phase chemical species \citep[see e.g.,][]{hase1992,garr2006,seme2010}. However, interstellar clouds are usually highly turbulent and permeated with magnetic fields \citep[e.g.,][]{aron1975,gold1995}. The dust grain motions in the clouds are strongly affected by the magnetohydrodynamic (MHD) turbulence through both hydrodrag and 2nd order Fermi acceleration \citep{Laza2002,yan2004, Yan2009, Hoang2012}. In particular, the gyroresonance mechanism proposed by \cite{YL2003} can accelerate the grains to supersonic speed relative to the gas. The turbulence induced grain motion has been implemented to the evolution modeling of interstellar grains \citep{hira2009}. In this work, it will be further considered in the gas-grain chemical modeling.

This paper is organised as follows: The physical and chemical models will be described in Sect.~\ref{models} as the basis of this work. We also introduce in this section the photoelectron ejection effect and a more proper treatment of the accretion of ions onto grains that have been added into our chemical model. The grain motion effect will then be integrated into grain accretion rates in Sect.~\ref{grainrole}. Then we present in Sect.~\ref{tools} several tools we will use in the analysis of the grain motion effect: timescales, age-normalised timescale plots and reaction rate tracing (RRT) diagrams. The detailed investigation of the chemical effects of grain motion is performed on the basis of detailed chemical modeling in three typical cloud environments (dark cloud, molecular cloud and cold neutral medium) in Sect.~\ref{results}. Finally, the most important findings are summarised in Sect.~\ref{summary}.

\section{Physical and chemical models}
\label{models}

\subsection{Physical models}
\label{phymodel}

The analysis of the grain motion effect in this work will be largely based on numerical gas-grain chemical simulations. As the basis of the chemical modeling, we adopt three physical cloud models: Dark Cloud (DC), Molecular Cloud (MC), and Cold Neutral Medium (CNM). They are typical interstellar medium models in which compressible magnetohydrodynamic turbulence and dust velocity relative to gas cannot be omitted \citep[$10^4-10^5$ cm\,s$^{-1}$;][]{yan2004}. To model the chemistry in these interstellar environments, we assume that the dust grains have a uniform grain radius $r_{\rm d}=10^{-5}$\,cm which is a typical value used in gas-grain models. The average turbulent grain speeds for grains of this radius in the three cloud models are interpolated from the computed data from \citet{yan2004} and listed in Table~\ref{tab_phy_models}. We do not consider the grain motion resulting from charge fluctuations as suggested by \citet{Ivle2010}. This is because we focus only on large grains. This effect is important for small grains and will be accounted for in future work including a distribution of different sizes of grains.

We compute six chemical models to test the effects of grain motion,
three standard models without grain motion (denoted as DC, MC and CNM)
and three models with grain motion (denoted as DCv, MCv and CNMv in Table~\ref{tab_phy_models}).
Most of the physical parameters of the models are also listed in the table. For the grain temperature,
as pointed out by \citet{math1983} and \citet{krue1984}, grains of different sizes exposed under different
 interstellar radiation strengths will have different temperatures. \citet{math1983} also showed that
 graphite grains can have higher temperature than silicates. In their radiation transfer computation
 of diffuse clouds and giant molecular clouds, the silicate grain temperature can vary from about $7-9$\,K
 at high $A_{\rm V}$ to $10-13$\,K at $A_{\rm V}=0$, while the graphite grain temperature can vary from $9-10$\,K
 at high $A_{\rm V}$ to $19-25$\,K at $A_{\rm V}=0$. For simplicity of this work, we will adopt a single
 representative grain temperature for each cloud model. For the DC model, it is usually acceptable to assume
 the same grain temperature $T_{\rm d}=10$\,K as the gas temperature. \citet{boul1996} measured dust temperatures
 of about 17.5\,K for high Galactic latitude HI clouds. The computed equilibrium dust temperatures for silicates
 and graphite grains by \citet{Li2001} also lie in the 16-20\,K range for a grain of size $0.1\micron$ under
 standard interstellar radiation field. Thus, we adopt $T_d=18$\,K for our CNM model, which is much lower than
 the gas temperature of 100\,K. For the MC model, the grain temperature is also very possibly slightly lower
 than the gas temperature. \citet{lada1981} obtained a grain temperature of about $10\pm 5$\,K in the molecular
 cloud B35 while its gas temperature was $23.4\pm 3.5$\,K; \citet{sarg1983} found a low dust temperature of about 13\,K
 and a high gas temperature of more than 30\,K in the dark cloud $\rho$\, Oph; \citet{gues1981} found a quite uniform
 warm gas temperature of $50-120$\,K near the Galactic center while the dust temperature is lower than 40\,K.
 Therefore, we adopt a representative grain temperature of $T_d=15$\,K for the MC model in this work,
 which is slightly lower than the adopted gas temperature of 25\,K. The visual extinction $A_{\rm V}$
 is set to values equivalent to the scaling of the interstellar radiation field in the work of \citet{yan2004}:
 $A_{\rm V}=7.5$ for the DC models,  2.5 for the MC model and 0.0 for the CNM model. The unattenuated far
 ultraviolet (FUV) flux $\chi$ is fixed to the standard Habing field $\chi_0$ for all models.
\begin{table}
 \centering
 \begin{minipage}{80mm}  
  \caption{Parameters for physical cloud models.}
  \label{tab_phy_models}
  \begin{tabular}{l@{\,}lll@{\,}ll@{\,}l}
  \hline
  \hline
   Model & $T$ & $T_d$ & $n_{\rm H}$   & $V_{\rm d}$       & $A_{\rm V}$   & $\chi$ \\
         & (K) & (K)   & (cm$^{-3}$)  & (km s$^{-1}$)    & (mag)         & ($\chi_0$) \\
    (1)  & (2) &  (3)  & (4)         & (5)              &  (6)          & (7) \\
   \hline
   DC    &  10 & 10 & $10^4$      & 0     & 7.5           & 1.0         \\
   DCv   &  10 & 10 & $10^4$      & 0.69  & 7.5           & 1.0         \\
   MC    &  25 & 15 & 300         & 0     & 2.5           & 1.0         \\
   MCv   &  25 & 15 & 300         & 0.49  & 2.5           & 1.0         \\
   CNM   & 100 & 18 & 30          & 0     & 0.0           & 1.0         \\
   CNMv  & 100 & 18 & 30          & 1.20  & 0.0           & 1.0         \\
\hline
\end{tabular} \\ [1mm]
The columns are:
(1) model name;
(2) gas kinetic temperature;
(3) grain temperature;
(4) gas density;
(5) average grain speed;
(6) visual extinction;
(7) unattenuated FUV flux expressed in units of the FUV interstellar radiation filed $\chi_0$ of \citet{drai1978}.\\
Notes:
The values of $T$, $n_{\rm H}$, and $V_{\rm d}$ are taken from \citet{yan2004}, the grain temperature $T_d$ is adopted from literature (see details in Sect.~\ref{phymodel}), $\chi$ is fixed to the standard Habing field in the interstellar conditions in solar neighborhood, $A_{\rm V}$ is fixed to the same values as in \citet{yan2004}.\\
\end{minipage}
\end{table}

\subsection{Chemical model}
We use a new gas-grain chemical code `ggchem' which is developed with reference to the code used in \citet[][]{hase1992}. This code is written in FORTRAN 90 language and, instead of the old ordinary deferential equation (ODE) solver (the gear's method), a new ODE solver (DVODE, using variable-coefficient method) from the public DOVDPK ODE package is used. The ggchem code has been successfully benchmarked using standard models from \citet{seme2010}. In this code, we follow the usual nomenclature rule to differentiate surface species from gas-phase species by prefixing the surface species by a letter 'J'. Thus, the same expressions such as `JCO' is used for surface species also in this paper.

The abundance of species $i$ will be denoted as $X(i)$ and always defined with respect to total H nuclei density $n_{\rm H}$ in this work.  The initial abundances of species for the DC model are listed in Table~\ref{tab_initabg87}, while that for the MC and CNM models are nearly the same except that the initial form of hydrogen is fully atomic in the latter two models.

The reaction network is based on the one used in \citet{seme2010} \footnote{It is available from the KIDA database: http://kida.obs.u-bordeaux1.fr/models}. It originally includes 655 species and 6067 reactions. Basically, the network contains gas-phase reactions, gas-grain interactions (accretion, grain charging processes, thermal desorption and cosmic ray induced desorption) and dust surface reactions. We have improved the network by introducing a more proper treatment of accretion of ions onto grains and adding photo-electron ejection (see details below), which  may be important for the MC and CNM models that are exposed to stronger interstellar radiation field. Particularly, the average grain charge will become positive in the MC and CNM models, so that the original network only with neutral grains and grains with one negative charge becomes insufficient. We have extended the grain charges to a wider range of -5 to +99. The average grain charge, together with ionization fraction, are the key parameters in determining the average grain drift speed in \citet{yan2004}. We set the cosmic ray ionisation rate $\zeta=1.3\times 10^{-17}$ for the DC model, $\zeta=2.0\times 10^{-16}$ for the MC model, and $\zeta=8.0\times 10^{-16}$ for the CNM model, so that the average grain charges and electron densities found in our chemical modeling are roughly consistent with that adopted by \citet{yan2004}. These choices of $\zeta$ are within the reasonable range determined from observations \citep[e.g.,][]{Lepp1992,Mcc2003,dalg2006,Ind2007,Ind2012,Vaup2014}.
\begin{table}
 \centering
 \begin{minipage}{80mm}
  \caption{Initial fractional abundances for DC model.}
  \label{tab_initabg87}
  \begin{tabular}{llll}
  \hline
  \hline
   species & $n(i)/n_{\rm H}$ $^*$ & species & $n(i)/n_{\rm H}$ $^*$ \\
   \hline
   H$_2^{**}$  & 5.0(-1) & Si$^+$ & 8.0(-9) \\
   He     & 9.0(-2)      & Fe$^+$ & 3.0(-9) \\
   C$^+$  & 1.2(-4)      & Na$^+$ & 2.0(-9) \\
   N      & 7.6(-5)      & Mg$^+$ & 7.0(-9) \\
   O      & 2.56(-4)     & P$^+$  & 2.0(-10)\\
   S$^+$  & 8.0(-8)      & CL$^+$ & 1.0(-9) \\
\hline
\end{tabular} \\ [1mm]
Note: Values are taken from \citet{seme2010}.\\
{\em *} a(b)=$a\times 10^b$.\\
{\em **} The only difference in MC and CNM models is that hydrogen is in atomic form and the initial abundance becomes unity.
\end{minipage}
\end{table}

In the following two subsections, we will give more details about the two grain charging processes (photo-electron emission and accretion of charges onto grains) which we have added to the chemical network.

\subsubsection{Photo-electron ejection}
\label{PE_effect}

The PE ejection from grains was not considered in the original
chemical network, which makes the network not suitable for the
models with lower dust extinction (MC and CNM models) in this work. We adopt the formulation initially developed by \citet{drai1978} and later improved by \citet{Wei2001} for the PE ejection effect. For simplicity, only silicate dust is considered in this work. For a given grain of radius $r_{\rm d}$ and charge $Z$, the photo-electron emission rate per grain can be obtained by integrating across interstellar radiation field frequency as
\begin{equation}
\label{JPE}
\begin{split}
  J_{PE}(Z,r_{\rm d})=\pi r_{\rm d}^2
   &\int_{\nu_{\rm pet}}^{\nu_{\rm max}}d\nu Y(\nu,Z,r_{\rm d}) Q_{\rm abs}(\nu,r_{\rm d})\frac{cu_\nu}{h\nu}\\
  +&\int_{\nu_{\rm pdt}}^{\nu_{\rm max}}d\nu \sigma_{\rm pdt}(\nu,Z,r_{\rm d})\frac{cu_\nu}{h\nu}.
\end{split}
\end{equation}
The first term at the right side of the equation is the rate for the photoionisation of valence electrons, while the second term is the photo-detachment rate for `extra' attached electrons on negatively charged grains. The PE yield $Y(\nu,Z,r_{\rm d})$ comprises three factors: the probability to produce a PE after the absorption of a UV photon, the probability for a PE to travel out of the bulk of the grain, and the probability for the PE to escape from the grain surface to infinity. The absorption coefficient $Q_{\rm abs}(\nu,r_{\rm d})$ is computed by Mie theory from given dielectric function of the grain material and spherical grain size $r_{\rm d}$ according to \citet{Li2001}; we used a FORTRAN code and dielectric function data file from Dr. Aigen Li through private communication. The cross section of the photo-detachment $\sigma_{\rm pdt}(\nu,Z,r_{\rm d})$ is also computed by a simple empirical formula from \citet{Wei2001}. The frequency integration limits are set by the availability of the most energetic photons in typical HI clouds ($h\nu_{\rm max}=13.6$\,eV) and the minimum energy required to knock out the PE: $h\nu_{\rm pet}$ for valence electrons and $h\nu_{\rm pdt}$ for loosely attached `extra' electrons only on negatively charged grains. See the full details in \citet{Wei2001}.

Then, the PE ejection rate coefficients can be computed according to
\begin{equation}
\label{KPE}
  k_{\rm PE}(Z,r_{\rm d})=J_{PE}(Z,r_{\rm d}) \times 10^{-A_{\rm V}/2.5}(\chi/\chi_0),
\end{equation}
where $\chi_0$ is the standard Habing field strength in UV (6-13.6\,eV). The PE ejection rate is $R_{\rm PE}(Z,r_{\rm d})=k_{\rm PE}(Z,r_{\rm d})n_d(Z,r_{\rm d})$.

\subsubsection{Accretion of charges onto grains}
\label{charging}

\begin{figure}

\centering
\fbox{
\includegraphics[scale=0.7]{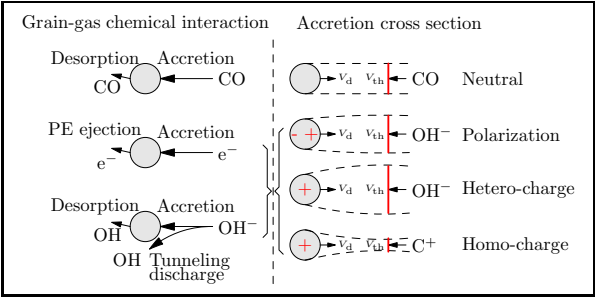}
}
\caption{Schematic for gas-grain interaction (left) and grain accretion cross section (right). The gray filled cycles are dust grains and the arrows indicate motions of particles. The left half shows the three typical cases of gas-grain interaction with neutral species, electrons and ions. The `Tunneling discharge' branching means ions can neutralise with oppositely charged grains through quantum tunneling before touching the grain surface and thus have a probability for their neutralised products to directly return back into gas phase. The right panel indicates the differences in accretion cross sections by the lengths of the red lines (not in proportion, however) for the four typical cases of accretion: neutral species onto all grains, ions onto neutral grains, ions onto hetero-charge grains and ions onto homo-charge grains. The details are discussed in section~\ref{charging}.}
\label{cartoon}
\end{figure}

Beside the missing of PE process, the original chemical network is also incomplete in the treatment of charge accretion onto grains, because only neutral grains and grains with one negative charge were considered. Moreover, the grain charging processes were treated in the original chemical network like normal chemical reactions with fixed reaction rate coefficients that are suitable only for a fixed grain charging condition, which makes it impossible to apply to models with different grain charging conditions and difficult to integrate the grain motion effects into the processes.

A more complete treatment of grain charging can be first divided into two broad categories: interactions of grains with electrons and with ions. For electrons, the PE process is the only electron ejection mechanism considered for grains in this work, while the opposite process of PE is the electron attachment to both neutral and charged grains. Ions can be accreted onto both neutral and charged grains, but the opposite process (ejection of ions from grains) is assumed to be impossible. This difference is visualised in the left half of the cartoon in Fig~\ref{cartoon}. However, ions and electrons are connected to each other by another pair of opposite processes: ionisation (by both UV photons and cosmic rays) and recombination.

The electron attachment cross section can be handled the same way as for the ion accretion process below.

The ion accretion process can be further divided into three sub-categories: accretion of ions onto neutral, hetero charge and homo charge grains. The ion accretion cross section, if compared with the geometrical cross section of the grains, will be slightly enhanced for neutral grains due to the polarisability of the grains, be significantly enhanced for hetero charge grains due to Coulomb attraction, and be significantly reduced for homo charge grains due to Coulomb repulsion, (as indicated in the right half of the cartoon in Fig~\ref{cartoon}). The detailed formulas for the charge accretion cross sections (applicable to both ions and electrons) will be given in detail below.

It is worth noting that the accretion of ions onto grains is also a process of material accretion which has similar chemical consequences as the accretion of neutral species onto grains. We assume that, together with their charges, ions are also accreted onto the neutral or homo charge grain surfaces, while for the case of ion accretion onto hetero charge grains, there is a probability for the neutralised products to return back into the gas phase instead of being accreted. The latter case is possible because the exchange of an electron can occur through quantum tunneling between the oppositely charged ion and grain without the requirement for the impinging ions to touch the grain surface. However, the sticking probability $p_{\rm neu}$ of the neutral products is unknown. We have tested two extreme cases: $p_{\rm neu}=0$ for the case in which all neutral products escape back to gas phase (this is also the case adopted in the original chemical network) and $p_{\rm neu}=1$ for the case in which all neutral products are adsorbed on the grain surfaces. It turns out that the different choices of $p_{\rm neu}$ have negligible effect in the CNM model and have a little effect only to the abundances of some surface species in the DC and MC models, but our conclusions upon the grain motion effect are not affected. Thus, we adopt the same choice of $p_{\rm neu}=0$ as in the original chemical network.

Also note that, for multi-atom ions whose neutral counterparts are unstable (e.g. H$_3^+$), they may break up into smaller stabler pieces after their charges are neutralised with hetero charge grains. This effect had been considered in the original chemical network for grains with one negative charge. What we have done is simply copying these reactions for grains with more negative charges. For the accretion of anions by positively charged grains, only 6 major simple anions are considered and their neutral counterparts are all stable (no need to break into smaller parts).

We utilise the formulas of \citet{drai1987} to describe the charge accretion cross section that is determined by Coulomb force and polarisability of the grains. As in that work, we first define a charge ratio $\nu=Ze/q_i$ ($Z$ is the number of charges on the grain, $e$ is the proton charge, $q_i$ is the charge of ionic species $i$) and thermal-to-electrical potential energy ratio $\tau=r_d k_{\rm b}T/q_i^2$ ($r_d$ is the grain radius, $k_{\rm b}$ is the Boltzmann constant, and $T$ is the kinetic temperature). The modification to the charge accretion cross section can be described by a Coulomb factor $\tilde{J}$ so that the cross section can be written as $\pi r_{\rm d}^2 \tilde{J}$. The formulas of $\tilde{J}$ can be found in \citet{drai1987} and we reproduce them for the three sub-categories below.

For the first sub-category (charge accretion onto a neutral grain), we have $\nu=0$, and the Coulomb factor averaged over thermal velocity is
\begin{equation}
\label{Jtilde_case1}
  \tilde{J}= 1+\left ( \frac{\pi}{2\tau} \right )^{1/2},
\end{equation}
which is different from unity merely due to the polarisability of the grain. After the charge accretion, the impinging species becomes a surface species (e.g., C$^+$ becomes JC).

For the second sub-category (charge neutralisation between an ion or electron and a hetero charge grain), we have $\nu=-|Z|<0$. Therefore, the Coulomb factor averaged over thermal velocity is
\begin{equation}
\label{Jtilde_case2}
  \tilde{J}\approx \left [1-\frac{\nu}{\tau} \right ] \left [ 1+\left ( \frac{2}{\tau-2\nu} \right )^{1/2} \right ].
\end{equation}
Here the factor in the first square brackets on the right side is due to Coulomb force, while the factor in the second square brackets is due to polarisability of the grain.

For the third sub-category (ion or electron accretion onto a homo charge grain), we have $\nu=|Z|>0$. Therefore, the Coulomb factor averaged over thermal velocity is
\begin{equation}
\label{Jtilde_case3}
  \tilde{J}\approx \left [1+(4\tau+3\nu)^{-1/2} \right ]^2 e^{-\theta_\nu/\tau},
\end{equation}
where $\theta_\nu \approx \nu/(1+\nu^{-1/2})$ is a dimensionless factor that represents the strength of the maximum Coulomb repulsive potential between the charged impinging particle and grain. Clearly, this process will be important only when the grain charge is not very large (small Coulomb repulsive potential) and the gas temperature is high enough (high kinetic energy of the impinging particle).

Then, the velocity averaged charge accretion rate coefficient for an ionic species $i$ is
\begin{equation}
\label{K_cha}
  K_{{\rm cha},i}(Z)=s_i\left ( \frac{8 k_{\rm b} T}{\pi m_i m_p} \right )^{1/2} \pi r_{\rm d}^2 \tilde{J},
\end{equation}
where $s_i$ and $m_i$ are the sticking coefficient of charge (onto grains) and mass number of the ion (or electron) respectively, $m_p$ is the proton mass. The value of $s_i$ is between $0\sim 1$; we adopt a fixed value of 0.5 in this work. It is convenient to average the charge accretion rate coefficient over all possible grain charges
\begin{equation}
\label{K_cha_ave}
  \overline{K_{{\rm cha},i}}=\sum_Z K_{{\rm cha},i}(Z)n_d(Z)/n_d,
\end{equation}
so that the total charge accretion rate between the ionic species $i$ and an average grain is
\begin{equation}
\label{R_cha}
  R_{\rm cha}(i)=\overline{K_{{\rm cha},i}}n(i)n_d,
\end{equation}
where $n(i)$ and $n_d(Z)$ are the number densities of the ion (or electron) and the grain with charge $Z$, respectively.

Taking the grain radius of $r_{\rm d}=10^{-5}$\,cm as an example, we obtain $\tau\approx 0.06$, $0.15$ and $0.60$, and thus $\tilde{J}\approx 6.12$, $4.24$ and $2.62$ for the first sub-category of charge accretion (accretion of charges onto neutral grains) at the temperatures of the DC, MC and CNM models, respectively. Therefore the accretion cross section is enhanced due to the grain polarisability only by factors of several in all three cloud models.

The representative grain charges in the three cloud models are found from our simulations to be close to $-e$, $+e$ and $+20e$ in DC, MC and CNM models, respectively, the representative values of $\nu$ are thus $\pm 1$ and $\pm20$. For the second sub-category of charge accretion (attractive charge accretion onto hetero charge grains), it is found that $\tilde{J}\approx 35.07$, $15.06$ and $41.95$ for the representative grain charges in the DC, MC and CNM models, respectively. Hence the Coulomb attraction significantly enhances the charge accretion cross section in all three cloud models.

For the third sub-category of charge accretion (repulsive charge accretion onto homo charge grains), we find $\tilde{J}\approx 5.8\times 10^{-4}$, $0.083$ and $1.87\times 10^{-12}$ for DC, MC and CNM models respectively. We thus conclude that the repulsive charge accretion is only mildly important in our MC model. However, as we will see in later sections, this process will become important also in the CNM model when the grain motion is included to help overcome the repulsive Coulomb barrier.

\section{Adding grain motion}
\label{grainrole}

\subsection{Introduce grain speed into grain related coefficients}
\label{addspeed}

To take into account the grain motion in gas-grain chemical models, we assume that both the dust grain velocity $\overrightarrow{v_{\rm d}}$ (with an average velocity $V_{\rm d}$) and the thermal velocity $\overrightarrow{V_{\rm t}}$ of a gas-phase species (with an average velocity $V_{\rm t}$) obey Gaussian distributions. Thus the relative velocity $\overrightarrow{v} = \overrightarrow{v_{\rm d}} - \overrightarrow{v_{\rm t}}$ also obeys a Gaussian distribution with an average velocity of $V^2=V_{\rm d}^2+V_{\rm t}^2$.

\subsubsection{Accretion of neutral species}
\label{addspeed_neutral}

In gas-grain chemical models \citep[see e.g.,][]{hase1992, garr2006, garr2008, seme2010}, the accretion rate (cm$^{-3}$s$^{-1}$) and rate coefficient (cm$^{-6}$s$^{-1}$) of a gas-phase neutral species $i$ onto grain surfaces are given by
\begin{equation}
\label{Racc_neu}
  R_{\rm acc}(i)=K_{\rm acc}(i)n(i)=K_{\rm acc}(i)X(i)n_{\rm H},
\end{equation}
\begin{equation}
\label{Kacc_aVn}
K_{\rm acc}(i)=s_n a_{\rm d} V_{\rm imp}(i) n_{\rm d},
\end{equation}
Here $s_n=1$ is the sticking probability, $a_{\rm d}=\pi r_{\rm d}^{2}$ is the geometric cross section of a grain with a radius $r_{\rm d}$ and
\begin{equation}
\label{nd}
n_{\rm d}=\frac{n_{\rm H} m_{\rm p} m_{\rm dg} }{(4/3)\pi  r_{\rm d}^3 \rho_{\rm d}}
\end{equation}
is the number density of the grains in which $ m_{\rm dg}=0.01$ is dust-to-gas mass ratio and $\rho_{\rm d}=3$ g\,cm$^{-3}$ is the bulk density of a dust grain. This definition of accretion rate coefficient is convenient because usually the number density of grains are assumed to be fixed. We further define the total geometric cross section in a unit volume as
\begin{equation}
\label{atot}
a_{\rm tot} =a_{\rm d}n_{\rm d}=\frac{n_{\rm H}}{r_{\rm d}} \frac{3m_{\rm p} m_{\rm dg} }{4\rho_{\rm d}}
\end{equation}
which, for given grain properties and dust-to-gas mass ratio, is proportional to the gas density. Thus, the rate coefficient becomes
\begin{equation}
\label{Kacc_atot}
K_{\rm acc}(i)=s_n a_{\rm tot} V_{\rm imp}(i).
\end{equation}
The impact velocity $V_{\rm imp}(i)$ is usually taken as the average thermal speed of species $i$
\begin{equation}
\label{vaccs}
V_{\rm imp}(i)=V_{\rm t}(i)=\sqrt{8k_{\rm b}T/(\pi m_i m_p)}.
\end{equation}

When the stochastic grain motion in turbulent interstellar gas is taken into account, the impact velocity should be replaced by the average relative speed between the gas and grains
\begin{equation}
\label{vaccd}
   V_{\rm imp}(i)= \sqrt{V_{\rm d}^2 + V_{\rm t}(i)^2},
\end{equation}
where the average speed $V_{\rm d}$ of the grain motion relative to gas has been investigated in MHD turbulence by \citet{yan2004}. Thus the rate of accretion for neutral species $i$ becomes
\begin{equation}
  R_{\rm acc,v}(i) =s_n a_{\rm tot} X(i) n_{\rm H} \sqrt{V_{\rm d}^2 +\frac{8k_{\rm b}T}{\pi m_i m_p}}
\label{racc}
\end{equation}
where we use the additional subscript `v' to denote quantities that include the grain motion effect.

\subsubsection{Accretion of ionic species}
\label{addspeed_charge}

For ions or electrons, as discussed in Sect.~\ref{charging}, they can be accreted onto both neutral and charged grains. While the charge accretion rate coefficient and rate have been given in Eqs.~(\ref{K_cha}) and (\ref{R_cha}), we prefer rewriting the rate coefficient as
\begin{equation}
\label{Kacc_cha}
  K_{{\rm cha},i}(Z)=s_i \tilde{a}_{\rm d} V_{\rm imp}(i),
\end{equation}
where the effective charge accretion cross section $\tilde{a}_{\rm d}=\pi r_{\rm d}^2\tilde{J}$. Hence, beside the similar involvement of the grain motion speed in the relative impact velocity $V_{\rm imp}$ as in Eq.~(\ref{vaccd}), the Coulomb factor $\tilde{J}$ is also affected through the involvement of kinetic temperature in the parameter $\tau$ which now becomes $\tau_{\rm v}=r_{\rm d}(k_{\rm b}T+m_i m_p V_d^2/2)/q_i^2$. The general trend of involving grain motion is thus to increase the impact velocity $V_{\rm imp}$ but to decrease the ion/electron accretion cross section $\tilde{a}_{\rm d}$ for neutral and hetero charge grains and to increase $\tilde{a}_{\rm d}$ for homo charge grains. Detailed examples are given in the context of the three cloud models below.

\subsection{Grain motion effect on grain related coefficients in the three cloud models}
\label{Racc_example}

\subsubsection{Accretion of neutrals}
\label{Racc_example_neu}

\begin{figure*}
 \centering
\includegraphics[angle=0,scale=.8]{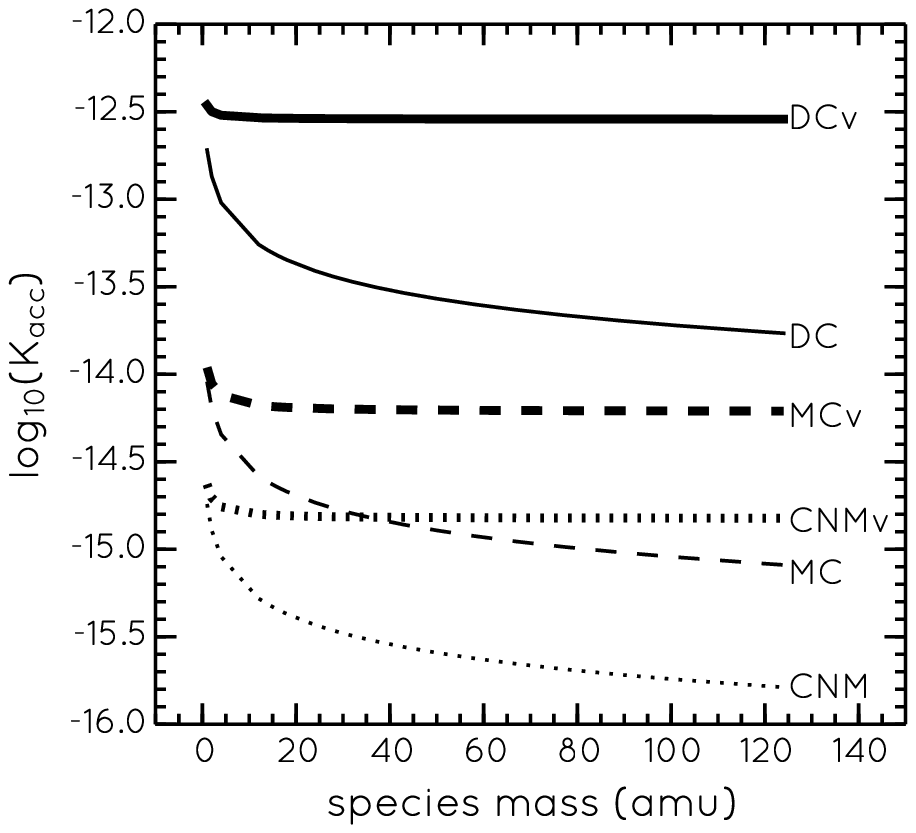}
\includegraphics[angle=0,scale=.8]{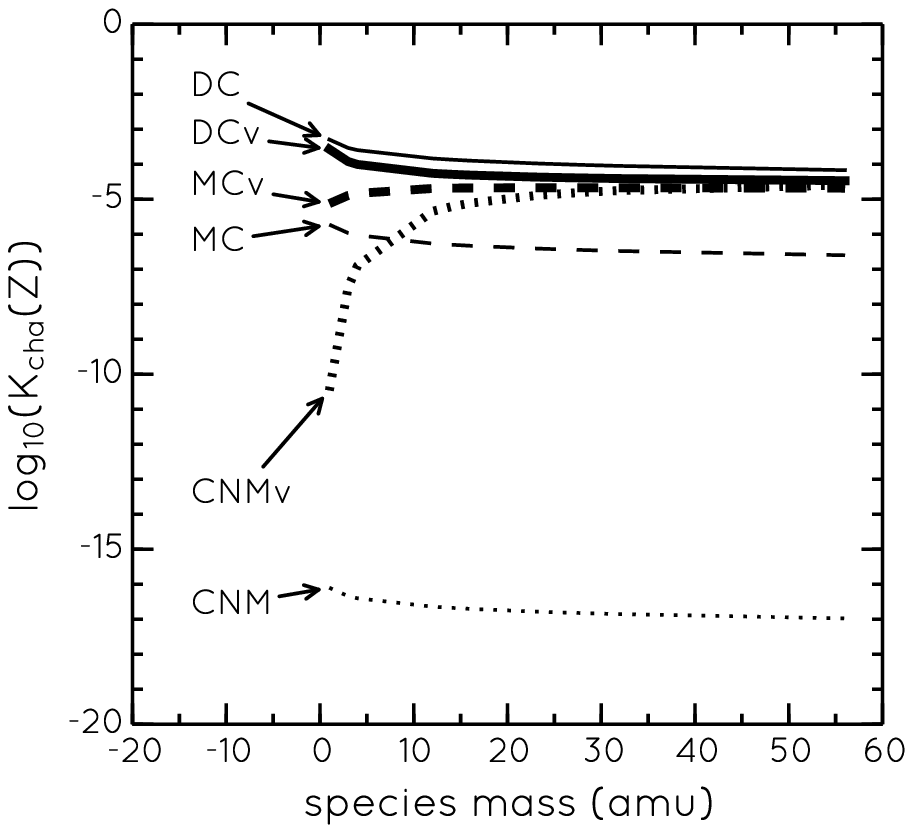}
 \caption{Effect of grain motion to grain accretion rate coefficients of neutral ({\em left panel}) and ionic ({\em right panel}) chemical species in the three molecular cloud models, DC (solid lines), MC (dashed lines) and CNM (dotted lines). The accretion rate coefficients with (thick lines) and without (thin lines) grain motion effect are plotted separately. Note that, in the right panel, only cations and electron (the left most point with nearly zero mass) are shown as examples. {\it (A color version of this figure is available online only.)} See the discussions in Sect.~\ref{Racc_example}.}
 \label{fig_kacc}
\end{figure*}

We illustrate the grain motion effects in Fig.~\ref{fig_kacc} by comparing the grain related rate coefficients between the models with and without grain motion in the three cloud environments, DC(v), MC(v) and CNM(v). For the neutral species in the left panel, the inclusion of grain motion greatly enhances the accretion of most neutral species other than the few lightest ones such as H, H$_2$ and He. Particularly, the uniform contribution of the grain motion to all species of various masses makes the accretion rate coefficients of most species (except the few lightest ones) almost constant, while the original coefficients ( without the grain motion effect) can vary by more than an order of magnitude among species. The enhancement of accretion coefficient is the largest for the heaviest species which amounts to factors of about 16, 7.9 and 9.4 in the DC, MC and CNM models respectively. The grain motion effect for the neutral species is the largest in the DC model because its lower gas temperature makes the grain accretion process more sensitive to the additional motion of the grains (see in Eqs.~\ref{atot}-\ref{vaccd}).

\subsubsection{Accretion of ions and electrons}
\label{Racc_example_ion}

For the accretion of ions in the right panel of the figure, the grain motion effect is found to be extremely sensitive to the number of charges on a grain. Here we adopt a representative grain charge of $-e$, $+e$ and $+20e$ as before for the DC, MC and CNM models respectively and only consider accretion of cations as an example. The inclusion of the grain motion is found to slightly reduce the charge neutralisation rate coefficients by a factor of several in the DC(v) models, but to inversely enhance the cation accretion rates onto homo charge grains by factors up to 100 in the MC(v) models, and to extremely boost up the cation accretion rates onto homo charge grains by huge factors up to more than $10^{12}$ in the CNM(v) models. The huge differences stem from the properties of the Coulomb interaction between the ions and the charged grains.

In the DC(v) models, the charge accretion is controlled by Coulomb attraction between the impinging cations and the negatively charged grains and thus the inclusion of the grain motion makes the Coulomb attraction more difficult.

Oppositely, in the MC(v) models, the Coulomb repulsion between the impinging cation and the `+e' charge on a grain impose a potential barrier to charge accretion. The kinetic energy of the impinging cations happens to be comparable to the barrier in the condition of the MC(v) models. Hence, the additional grain motion helps the impinging cations to overcome the Coulomb barrier more easily and the exponential dependence of the charge accretion rate on the ion-grain relative motion (see the involvement of $\nu$ in the quantity $\theta_\nu$ in Eq.~\ref{Jtilde_case3}) results in a large enhancement of the charge accretion rate coefficient in the MC(v) models.

The situation of the CNM(v) models is similar as in the case of MC(v), except the fact that the much higher grain charge of $+20e$ creates so high a Coulomb barrier that, without the grain motion, the cation thermal energy is significantly lower than barrier, which makes the cation accretion onto the $+20e$ charge grains almost impossible. The physical conditions in the CNM(v) models are in such a subtle state that the high average grain speed of 1.2\,km\,s$^{-1}$ in the CNMv model can help the impinging cations to gain a kinetic energy comparable or even higher than the Coulomb barrier, which results in the huge exponential rise up of the charge accretion rate coefficient observed in the right panel of Fig.~\ref{fig_kacc}.

Note that the inclusion of the grain motion generally also tends to increase the velocity factor $V_{\rm imp}(i)$ of the charge accretion rate (Eqs.~\ref{Kacc_atot}-\ref{vaccd}). But this factor only plays a minor role compared to the above Coulomb force factor.

In a summary, the general effect of grain motion is to speed up the accretion of neutral species onto grains, slow down ion-grain charge neutralisation, but exponentially boosts the accretion of ions onto homo charge grains. We did not exemplify the accretion of ions onto neutral grains. But it is easy to infer that it will be slightly reduced by the grain motion effect, just as the case of ion-grain charge neutralisation.

Furthermore, for the grain motion effect in the gas-grain interaction to take effect in the chemical models, one still needs to consider the evolution of the densities $n(i)$, $n(e)$ and $n_{\rm d}(Z)$ and to make sure these processes are not overwhelmed by other competing processes such as grain desorption and chemical reactions. Therefore, several tools such as timescale analysis and reaction rate tracing diagram are introduced in the next section to help understand when and how the grain motion takes effect.

\section{Tools for understanding the chemical effects of grain motion}
\label{tools}

\subsection{Definition of timescales}
\label{timescales}

In pseudo-time dependent chemical simulations, the importance of a chemical process to a given chemical species can be characterised by its time scale. Generally, one can differentiate two major categories of process for a given species $i$: its production and consumption processes. Assuming the corresponding reaction rates at a given time $t$ are $R_{\rm pro,tot}(i,t)$ and $R_{\rm con,tot}(i,t)$, respectively, and the current density of the species is $n(i,t)$, the production and consumption timescales can be defined as
\begin{equation}
\label{pro_timescales}
t_{\rm pro,tot}(i,t)=n(i,t)/R_{\rm pro,tot}(i,t),
\end{equation}
\begin{equation}
\label{con_timescales}
t_{\rm con,tot}(i,t)=n(i,t)/R_{\rm con,tot}(i,t).
\end{equation}
The meaning of timescales defined this way is that, for given ideally fixed current conditions, the current density of the considered species will be zeroed out (for consumption processes) or doubled (for production processes) within the defined timescale.

We separate the grain related processes (accretion of neutral and charged particles onto neutral and charged grains, desorption of neutral species from grain surfaces, and ejection of photo-electrons) from the chemical reactions and photo-processes in the gas and on the surfaces. Therefore, now we have four similarly defined timescales for any gas-phase or surface species $i$:
\begin{itemize}
\label{four_timescales}
\item{$t_{\rm acc}(i,t)$ -- the grain accretion timescale;}
\item{$t_{\rm des}(i,t)$ -- the grain desorption timescale;}
\item{$t_{\rm pro}(i,t)$ -- the chemical production timescale accounting for all chemical reactions that make the species;}
\item{$t_{\rm con}(i,t)$ -- the chemical consumption timescale accounting for all chemical reactions that destroy the species.}
\end{itemize}

The accretion timescale applies to accretion of neutral species, electron attachment and ion accretion processes, while the desorption timescale applies to both desorption of neutral species and ejection of photo-electrons from grains. The grain desorption process only includes thermal and cosmic ray induced desorption. The total desorption rate of a surface species $i$ is
\begin{equation}
\label{Rdes}
R_{\rm des}(i,t)=n(i,t)(K_{\rm td}(i)+K_{\rm cd}(i))
\end{equation}
in which the time-independent thermal and c.r. induced desorption rate coefficients are defined as
\begin{equation}
\label{Ktd}
K_{\rm td}(i)=\nu(i)exp(-\frac{E{\rm _d}(i)}{T_{\rm d}}),
\end{equation}
\begin{equation}
\label{Kcd}
K_{\rm cd}(i)=f\nu(i)exp(-\frac{E{\rm _d}(i)}{70 K}),
\end{equation}
where $\nu(i)=\sqrt{\frac{2N_{\rm s} k_{\rm b} E_{\rm d}(i)}{\pi^2 m_i m_p}}$ is the characteristic vibrational frequency of species $i$, $k_{\rm b}$ is Boltzmann constant in erg\,K$^{-1}$, $E_{\rm d}(i)$ is the desorption energy in unit of K, $N_{\rm s}=1.5\times 10^{15}$ (sites cm$^{-2}$) is the adsorption site density on grain surface, and $f=3.0\times 10^{-19}$ is the ratio of a grain cooling timescale via desorption of molecules to the timescale of subsequent heating events by c.r. \citep[e.g.][]{seme2010}.

For the gas-phase and surface chemical reactions, the chemical production and consumption rates are the sum of the individual rates of all reactions involving the consider species $i$. These rates are usually related to the evolving abundances of various related species and thus the two corresponding timescales, $t_{\rm pro}(i,t)$ and $t_{\rm con}(i,t)$, are generally time-dependent and show complex behavior during the model evolution. Their detailed properties will be analysed with a more powerful tool --- the reaction rate tracing (RRT) method --- in later sections of this work.

\subsection{Special properties of the accretion and desorption timescales}
\label{tacc_tdes}

Here we stress that the evolution of gas and surface abundances have different dependence on the accretion and desorption rate coefficients. For a gas-phase species $i$, the accretion timescale is the gas depletion timescale
\begin{equation}
\label{tacc_g}
t_{\rm dep}(i)=\frac{n_{\rm g}(i,t)}{R_{\rm acc}(i,t)}=\frac{1}{K_{\rm acc}(i)}
\end{equation}
from which one can see that it has no relation to its current density, but is only related to the grain properties and gas temperature, according to Eqs.~(\ref{Kacc_aVn}-\ref{vaccd}). Thus, this timescale does not change with time during the model cloud evolution. For a surface species $i$, the accretion timescale becomes the ice(-mantle) growth timescale
\begin{equation}
\label{tacc_s}
t_{\rm gro}(i,t)=\frac{n_{\rm s}(i,t)}{R_{\rm acc}(i,t)}=\frac{n_{\rm s}(i,t)}{n_{\rm g}(i,t)}t_{\rm dep}(i).
\end{equation}
This definition shows that, while the ice(-mantle) growth timescale is proportional to the time-independent depletion timescale $t_{\rm dep}(i)$ of its gas-phase counterpart, $t_{\rm gro}(i,t)$ itself becomes time-dependent due to the involvement of the evolving surface-to-gas density contrast.

Similarly, the desorption timescale for a surface species $i$ is the ice(-mantle) evaporation timescale
\begin{equation}
\label{tdes_s}
t_{\rm eva}(i)=\frac{n_{\rm s}(i,t)}{R_{\rm des}(i,t)}=\frac{1}{K_{\rm des}(i)},
\end{equation}
which is only related to the grain properties and is thus time-independent, according to Eqs.~(\ref{Rdes}-\ref{Kcd}). While the absorption timescale for the gas-phase species $i$ becomes the gas enrichment timescale by desorption
\begin{equation}
\label{tdes_g}
t_{\rm enr}(i,t)=\frac{n_{\rm g}(i,t)}{R_{\rm des}(i,t)}=\frac{n_{\rm g}(i,t)}{n_{\rm s}(i,t)}t_{\rm eva}(i).
\end{equation}
It is again related to the gas-to-surface density contrast of the considered species, which makes it time-dependent.
For ions, only one grain related timescale, the charge accretion timescale, can be defined as
\begin{equation}
\label{tneu_ion}
t_{\rm cha}(i,t) = \frac{n_g(i,t)}{R_{\rm cha}(i,t)} = \frac{1}{\sum_Z K_{{\rm cha},i}(Z)n_d(Z,t)}.
\end{equation}
It is time dependent because of the involvement of the evolving grain charge distribution. We do not discuss the timescales of electrons, because little grain motion effect is expected for them.

\subsection{Age-normalised timescale}
\label{grainspeed_importance}

During the evolution of chemical models, for any of the above four processes to be chemically important, its timescale must be comparable or shorter than the age of the model. This definition of importance can be understood as that, if the current density and rates are fixed, a chemically important process should be able to zero out or double the current density $n(i)$ when the age of the model doubles. Therefore, the chemical importance of the four processes can be better characterised by age-normalised timescales
\begin{equation}
\label{tnormpro}
\tilde{t}_{\rm pro}(i,t)=t_{\rm pro}(i,t)/t,
\end{equation}
\begin{equation}
\label{tnormcon}
\tilde{t}_{\rm con}(i,t)=t_{\rm con}(i,t)/t,
\end{equation}
\begin{equation}
\label{tnormdep}
\tilde{t}_{\rm dep}(i,t)=t_{\rm dep}(i,t)/t {\rm \hspace{4pt}(for\hspace{2pt} gas\hspace{2pt} species)},
\end{equation}
\begin{equation}
\label{tnormenr}
\tilde{t}_{\rm enr}(i,t)=t_{\rm enr}(i,t)/t {\rm \hspace{4pt}(for\hspace{2pt} gas\hspace{2pt} species)},
\end{equation}
\begin{equation}
\label{tnormgro}
\tilde{t}_{\rm gro}(i,t)=t_{\rm gro}(i,t)/t {\rm \hspace{4pt}(for\hspace{2pt} surface\hspace{2pt} species)},
\end{equation}
\begin{equation}
\label{tnormeva}
\tilde{t}_{\rm eva}(i,t)=t_{\rm eva}(i,t)/t {\rm \hspace{4pt}(for\hspace{2pt} surface\hspace{2pt} species)},
\end{equation}
\begin{equation}
\label{tnormneu_ion}
\tilde{t}_{\rm cha}(i,t) = t_{\rm cha}(i,t)/t {\rm (for\hspace{4pt}ions)},
\end{equation}
where $t$ is the model age. Because the grain motion enters the problem only through the accretion processes, its effect will be important only when the accretion processes are important (when $\tilde{t}_{\rm dep}(i,t)\leq 1$, $\tilde{t}_{\rm gro}(i,t)\leq 1$ or $\tilde{t}_{\rm cha}(i,t)\leq 1$).

When the accretion timescale is already shorter than the cloud age, the chemical importance of accretion (and of grain motion effect) further depends on whether it is not overwhelmed by other processes such as desorption, chemical production and consumption. The processes with the shortest age-normalised timescales dominate the abundance evolution of the considered species. The relative importance of the various processes usually evolves with time.

\subsection{Reaction rate tracing (RRT) method}
\label{RRT}

When the grain accretion process is at least not the only dominant process and the chemical reactions play a crucial role for a given species $i$, one may need to trace more details of the evolving network of chemical reactions to interpret the grain motion effects. This is particularly the case for most ionic species whose abundances are generally dominated by gas-phase chemical reactions in our model. In this case, we will apply the reaction rate tracing (RRT) method in which, in addition to the plots of abundance evolution tracks and age-normalised timescale curves define in previous subsections, plots of total production and consumption rates and rates of individual leading processes (RRT plots) will be considered. The chemical formulae of the leading reactions will be also listed out in the plots to help recognise them. Such a combination of the abundance, age-normalised timescale, and RRT plots is called an RRT diagram in this work.

The leading processes to be shown in the RRT plots are selected in the following two steps: First, we consider the rates of all processes at a given time $t$ and select the smallest number of the fastest processes whose collective contribution to the total production/consumption rate is no less than a specified threshold percentage $\alpha$. Second, this selection procedure is repeated for all evolution time steps of the chemical modeling and all the selections are merged to yield the final set of leading processes to show in the RRT plots. The threshold $\alpha$ can be arbitrarily adjusted between 0 and 1 to increase or reduce the number of processes to show in the RRT plots. When $\alpha=0$, only those reactions that are the fastest at least at one time step are selected; while when $\alpha=1$, all reactions will be selected. For the RRT plots shown in this work, we set $\alpha=0$.

In order to investigate the grain motion effect, we also add an additional row of sub-plots to the bottom of the RRT diagrams to compare the RRT plots between the models with and without the inclusion of grain motion. Examples of the RRT diagram can be found in the discussion of ionic species in the DC(v) models in Sect.~\ref{chem_example_ion}.

\section{Results}
\label{results}

This section summarises the overall properties of grain motion effects in the DC, MC and CNM models. Particularly, we will use individual examples in the DC(v) models to illustrate how and when the grain motion begins to take effect in chemistry. Then the grain motion effects are compared among the three cloud models and interpretation to the findings will be given.

\subsection{Dark cloud (DC) models}
\label{DC}

We have compared the abundance evolution tracks between the DC and DCv models for all species whose maximum number density is no less than $10^{-15}$\,cm$^{-3}$ to check the grain motion effects. It is found that most of both the gas-phase and surface species are definitely affected by the grain motion effect, with only few exceptions which will be mentioned in the next paragraphs. For the gas-phase species, no matter whether they are neutrals or ions, the grain motion effect manifests itself as two types of typical phenomena: 1) during certain periods of slow abundance evolution, the gas-phase abundances are lowered by 2-3 orders of magnitude due to the enhanced grain accretion by the moving grains; 2) at those special moments when the species experiences sharp changes in abundance, the inclusion of grain motion usually tends to make the sharp changes to occur at ages about ten times earlier.

Most of the surface ice species also show grain motion effects in their abundance evolution tracks. The effects can be grouped into two classes too: 1) at the beginning of the modeling or when the abundances do not vary sharply, the inclusion of grain motion can enhance the surface abundances by varying factors up to two orders of magnitude (e.g., the JOH, JHCN, JCN, JNO, JS, JP ices, etc.); 2) at the moments of drastic abundance jumps, the occurrence of the jumps is usually postponed to a later age by a factor of several by the grain motion effect (e.g., this is the case for the JNH, JNO, JOH, JCN JS, JH, JCH$_2$, and JCH$_3$ ices among others). Several special cases will be discussed below.

Some special cases exist for the gas-phase species. For example, the abundances of some ionic species such as C$^+$ and C$_2^+$ are very insensitive to the grain motion effect. This is because both the production and consumption reactions involve neutral species and are both reduced to similar factors by the grain motion effect, which leaves the considered abundance almost unchanged. Some other species such as O$_3$ and CH$_3$ and ionic species such as H$_3^+$, He$^+$, HeH$^+$, H$^+$ and CH$_5^+$ show enhanced abundance in some periods of their chemical evolution, which is counter intuitive. The enhancement of these ionic abundances is due to the decrease of their neutral destroyers by the grain motion effect, as discussed for the case of H$_3^+$ in Sect.~\ref{chem_example_ion}. For the neutral species CH$_3$, its abundance is enhanced, opposite to the majority of gas-phase neutral species, due to the fact that it is mainly made from the enriched ion CH$_5^+$. In the special case of O$_3$, its abundance is also enhanced by the grain motion effect, which is because it is mainly made through the enhanced surface reactions between JO atoms and subsequent desorption.

Some special cases for the surface species include the JMgH$_2$, JH$_2$O, JH$_2$S, JSiH$_4$ ices whose abundance seem to be insensitive to the inclusion of grain motion. The reason is the same as will be explained in the discussions of the JH$_2$O ice case in Sect.~\ref{chem_example_surface}: cancellation between the increase and decrease of abundances of two or more building blocks of the ices in the grain surface reactions. Two other species, JH and JNH$_3$ ices, unexpectedly show decrease of abundance by the grain motion effect, which is because JH atoms are heavily consumed by many other enriched neutral surface species through surface reactions and the formation of ammonia ice heavily relies on the availability of JH.

Below we will demonstrate how the grain motion alters the chemical abundance evolution tracks of some representative neutral, ionic and surface species in the DC(v) models.

\subsubsection{Gas-phase neutral species}
\label{chem_example_gas_neutral}

Before going to the detailed examples, we recall that, for a gas-phase neutral species, the grain accretion process is called depletion process while the desorption process is termed enrichment process. Therefore, we will intensively use the corresponding timescale $t_{\rm dep}$ and $t_{\rm enr}$ in the discussions below.

\begin{figure*}
\centering
\includegraphics[angle=0,scale=0.72 ]{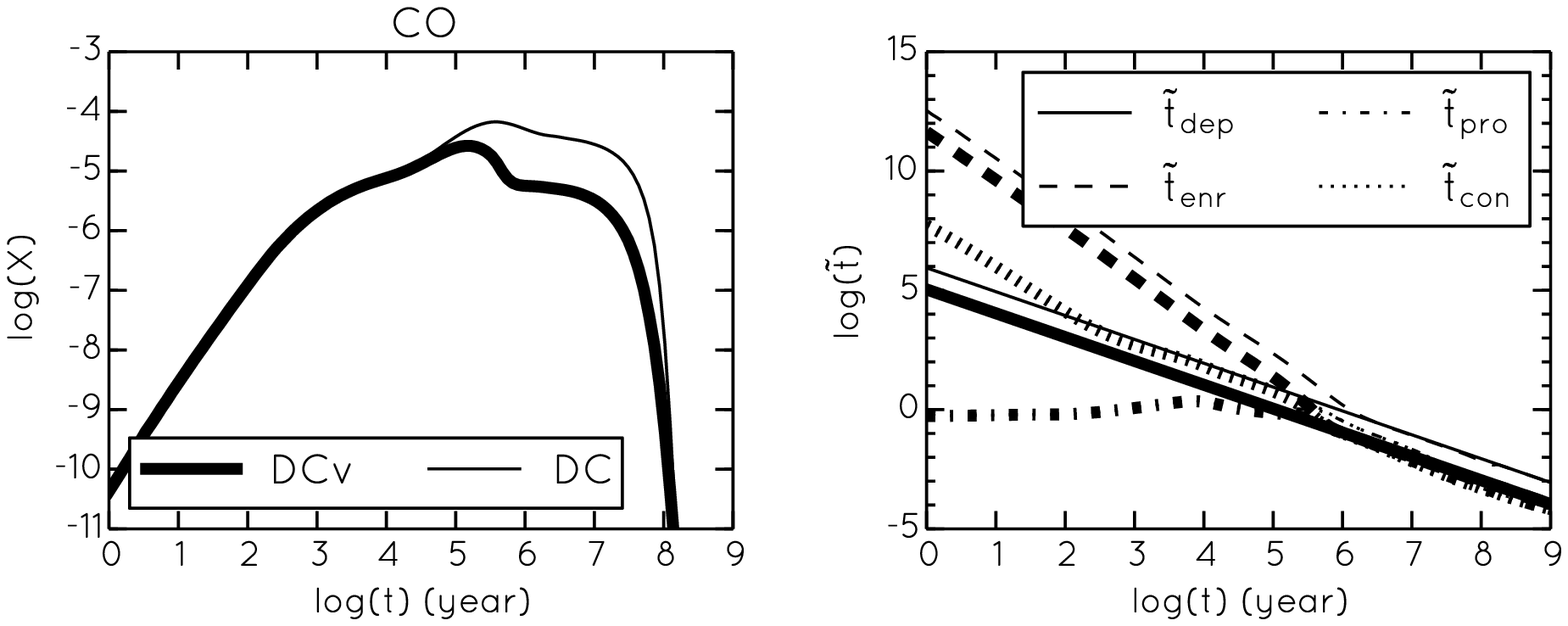}
\includegraphics[angle=0,scale=0.72 ]{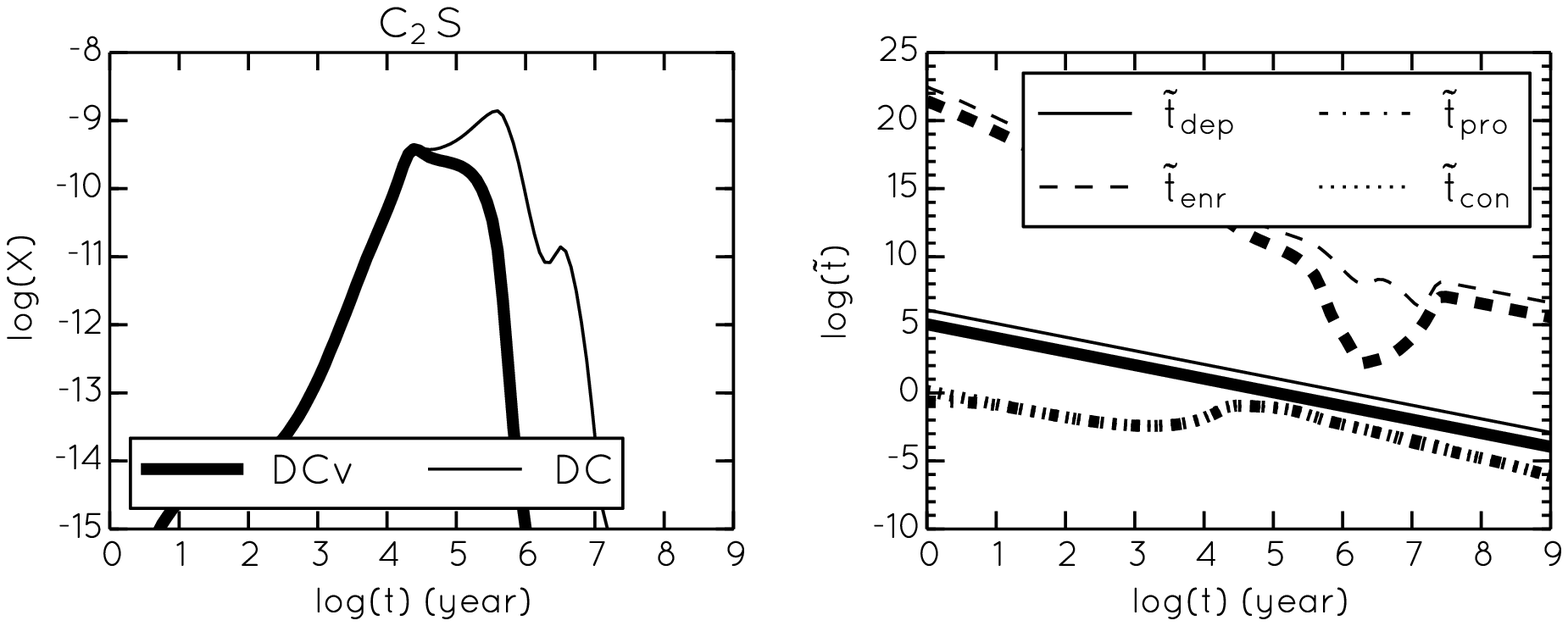}
\includegraphics[angle=0,scale=0.72 ]{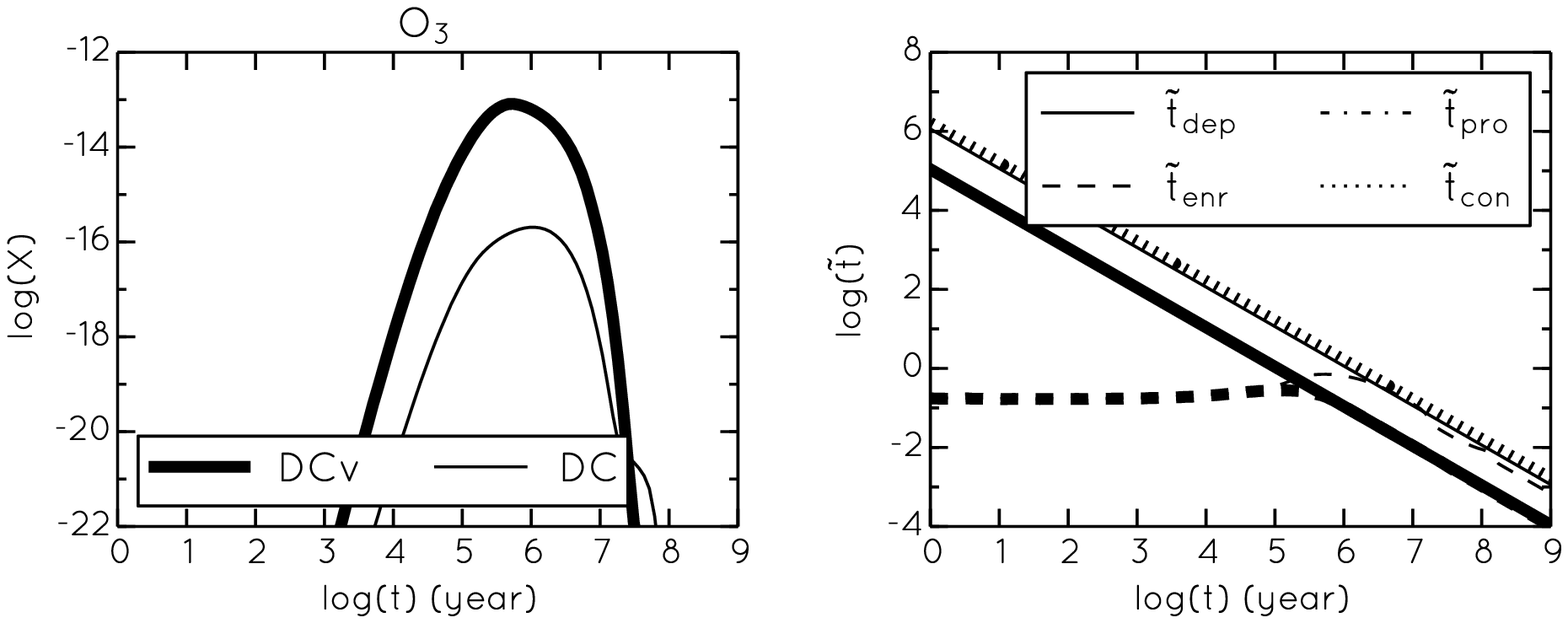}
\caption{Chemical effects of grain motion to gas-phase neutral species in three representative cases (one case per row) in the dark cloud (DC) model.  In each row, the left panel is the abundance evolution tracks for models with (thick black lines) and without (thin black lines) grain motion; the right panel is the corresponding age-normalised timescale curves of the four processes (gas depletion by solid line, gas enrichment by dashed line, chemical production by dash-dotted line, and chemical consumption by dotted line; the use of thick black and thin black lines to differentiate models with and without grain motion effects is the same as in the left panels) as defined in Eqs.~(\ref{tnormpro}-\ref{tnormenr}). See the discussions in Sect.~\ref{chem_example_gas_neutral}. }
\label{X_IM_neu}
\end{figure*}
{\noindent \it CO}\\
Three examples of neutral species will be discussed. They are presented as the three rows in Fig.~\ref{X_IM_neu}, with the top row for the first example species, CO. The abundance evolution tracks (top left panel) show that the inclusion of grain motion in the model will bring down the CO abundance by an order of magnitude only after an age of about $10^4-10^5$\,yr. The corresponding logarithmic age-normalised timescale plot (top right panel) demonstrates that, before the age of $10^4-10^5$\,yr, the gas-phase chemical production processes (dash-dotted lines) have the smallest age-normalised timescale and thus dominate over the chemical consumption and grain related processes. The grain accretion process is unimportant during this age because the accretion timescale of about $10^5$\,yr is much longer than the cloud age, which explains why the grain motion effect in CO does not appear until after this age. After the model cloud is older than about $10^5$\,yr, all four processes become comparably chemically important, as can be seen from the crowding of the age-normalised timescale curves in the figure. With the inclusion of grain motion, both grain accretion and desorption are enhanced (with smaller age-normalised timescales as shown by the thick black curves in the figure), while the two chemical reaction processes show little difference (the corresponding thick black dash-dotted and dotted curves are almost overlapped with the thin black curves in the figure). The enhancement of the desorption process (enrichment of CO) is an indirect effect of grain motion: it is caused by the increase of the surface JCO abundance before $10^5$\,yr and by the decrease of gas-phase CO abundance after then (see the definition of the enrichment timescale in Eq.~\ref{tdes_g}). Therefore, the grain motion effect in CO abundance is driven directly through the enhancement of accretion process and shows up mainly in the late period.

We stress here that the lower CO abundance due to grain motion effect may not
contradict with the generally found CO abundance of about $10^{-4}$ in molecular clouds,
because we did not fine tune our chemical and physical model parameters to meet the observations.
However, a more detailed investigation of the CO abundance in future may provide a valuable
observational test to the grain motions and their chemical effects.

{\noindent \it C$_2$S}\\
The evolution tracks (middle left panel) and age-normalised timescale curves (middle right panel) of C$_2$S are shown in the middle row of Fig.~\ref{X_IM_neu}. The grain motion effect similarly begins to emerge after a model cloud age of about $10^4-10^5$\,yr to bring down the gas-phase C$_2$S abundance. However, the grain motion effect in this case is an indirect effect. The chemistry of C$_2$S is always dominated by fast chemical production and consumption reactions, as can be seen from the overlapping dash-dotted and dotted lines in the middle right panel of the figure. The grain motion effect decreases its abundance at later ages by reducing the available amount of the raw chemical material (e.g., C and S) that makes C$_2$S. A large body of gas-phase neutral species also belong to this case.

{\noindent \it O$_3$}\\
The evolution tracks (bottom left panel) and age-normalised timescale curves (bottom right panel) of O$_3$ are shown in Fig.~\ref{X_IM_neu}. Its abundance is counter-intuitively greatly enhanced by the inclusion of grain motion from the very beginning of the chemical modeling. The age-normalised timescale plot shows that its abundance evolution is solely controlled by desorption process before the age of $10^5$\,yr, which means the gas-phase ozone should be mainly formed on grains and the enhancement of the gas-phase ozone abundance is a secondary effect of the enhanced abundances of JO atoms due to the inclusion of grain motion. This result is a natural consequence of our chemical network in which there is no gas-phase formation channel for O$_3$ (correspondingly, the curve for $\tilde{t}_{\rm pro}$ is missing in the bottom right panel of Fig.~\ref{X_IM_neu}).

\subsubsection{Surface species}
\label{chem_example_surface}

Again, we recall that, for a surface icy species, the grain accretion process is termed ice growth process, while the desorption process is called ice evaporation process. The corresponding times intensively used in this subsection will thus be $t_{\rm gro}$ and $t_{\rm eva}$.

\begin{figure*}
\centering
\includegraphics[angle=0,scale=0.72 ]{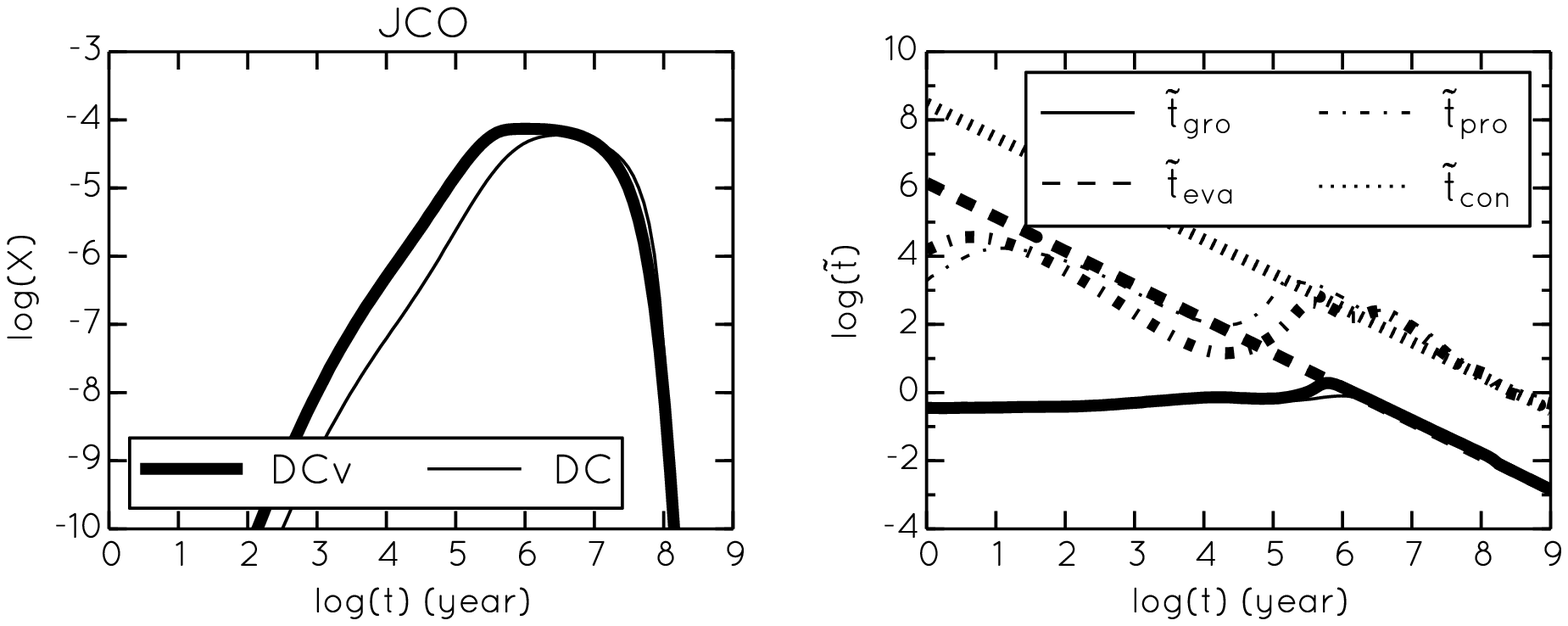}
\includegraphics[angle=0,scale=0.72 ]{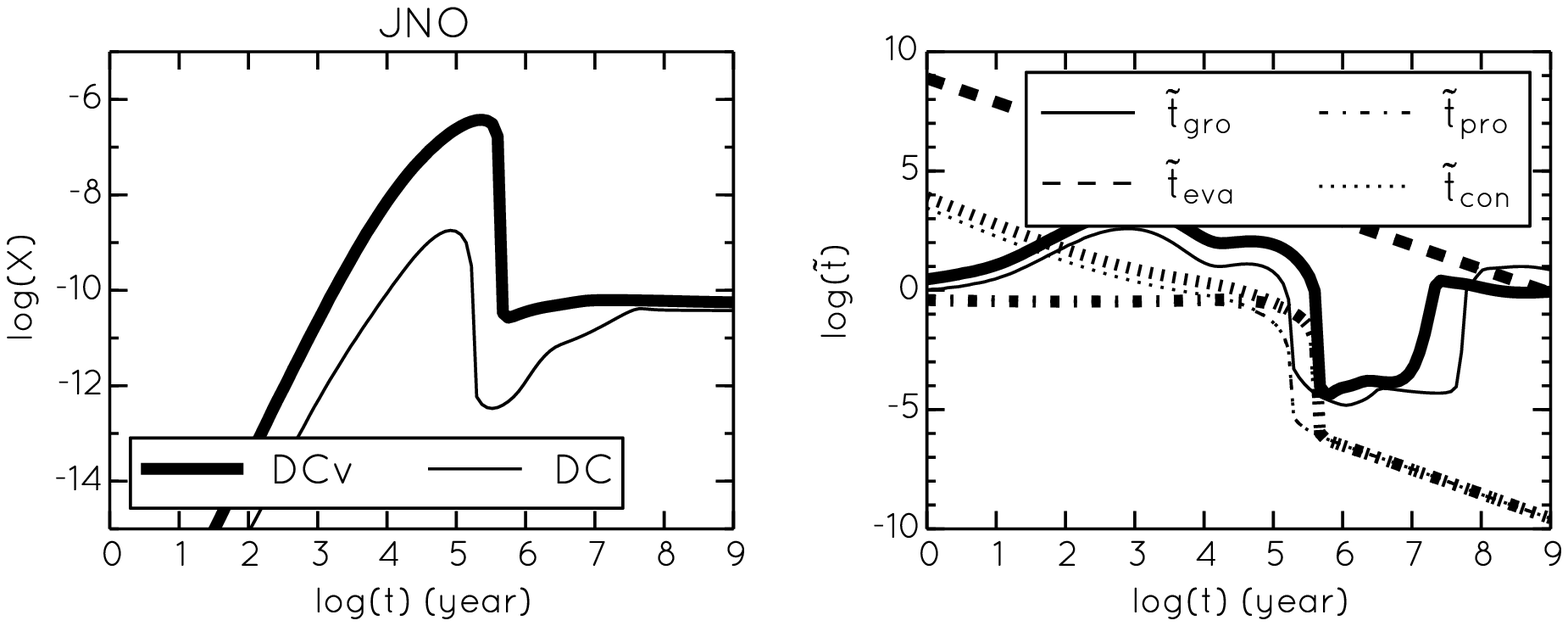}
\includegraphics[angle=0,scale=0.72 ]{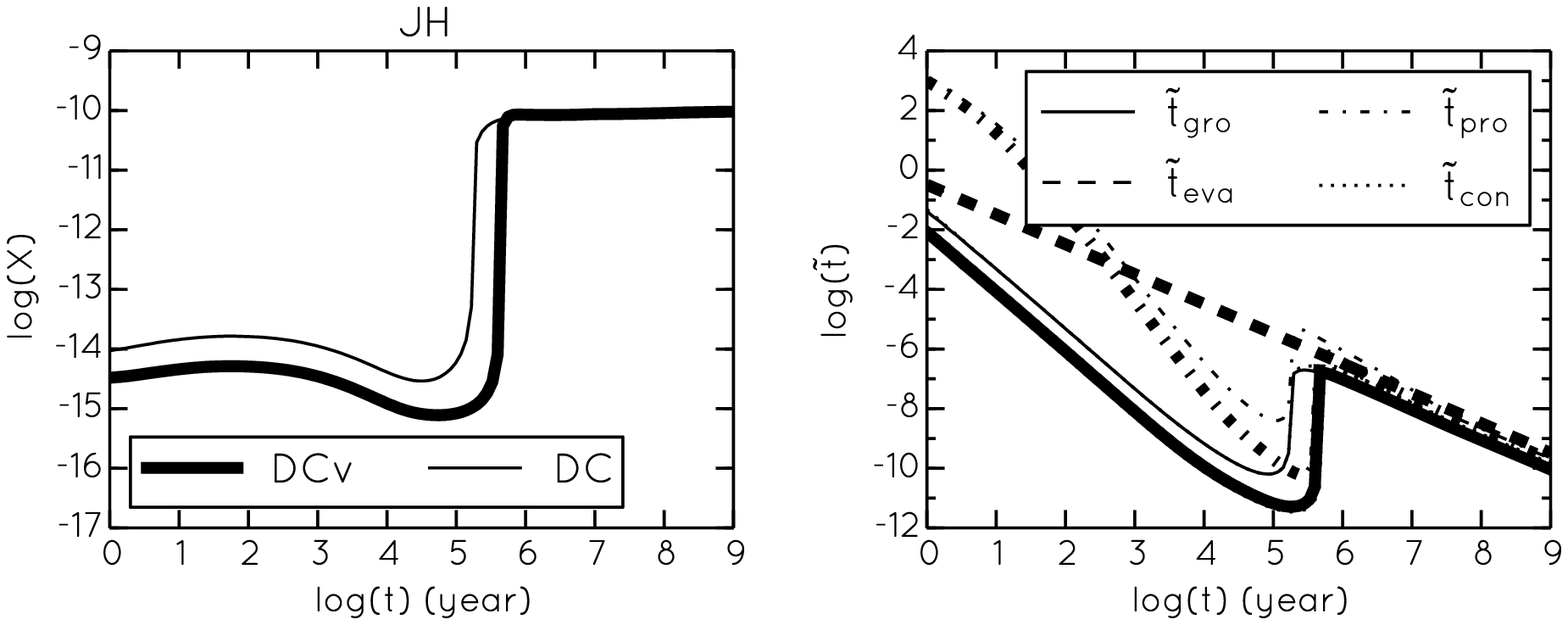}
\includegraphics[angle=0,scale=0.72 ]{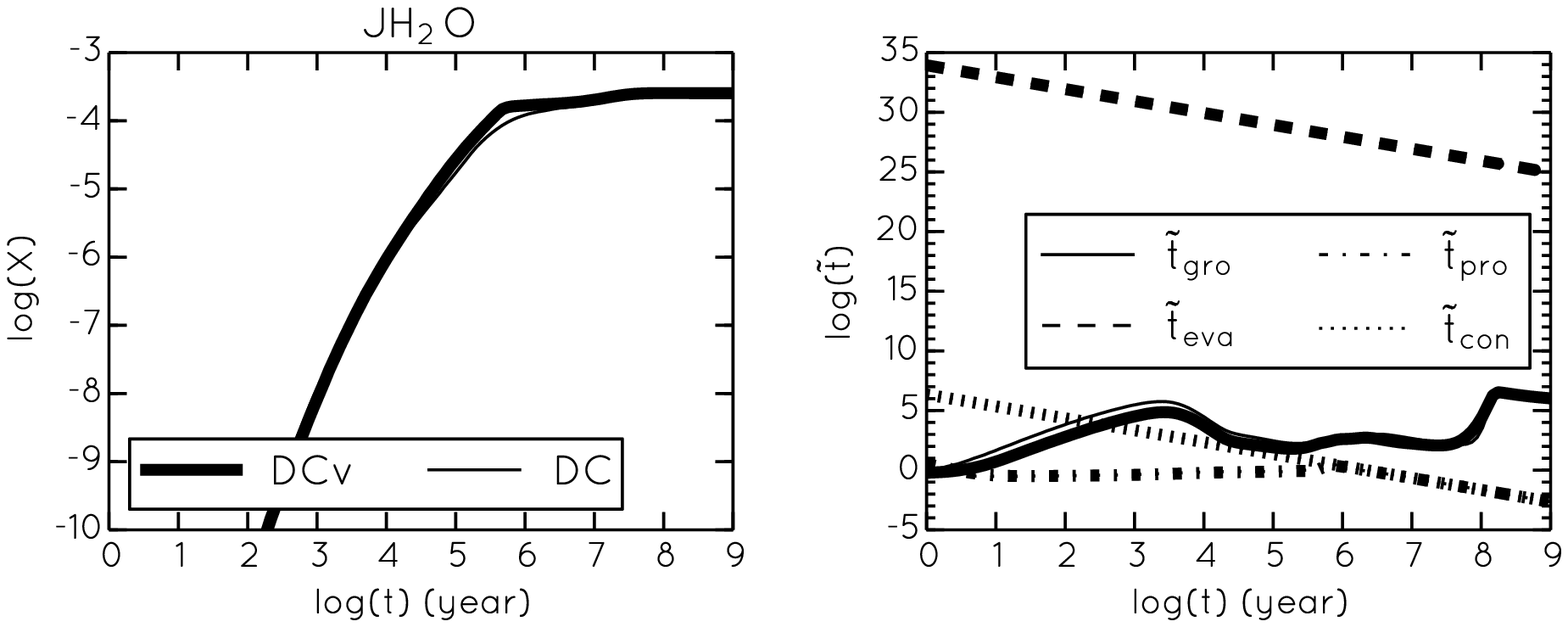}
\caption{Same as Fig.~\ref{X_IM_neu} but for representative surface ice species. The age-normalised timescales (ice growth timescale, ice evaporation timescale, and surface chemical production and consumption timescales) in the right panels are defined in Eqs.~(\ref{tnormpro}, \ref{tnormcon}, \ref{tnormgro}, \ref{tnormeva}). See the discussions in Sect.~\ref{chem_example_surface}. }
\label{JX_IM_surf}
\end{figure*}
{\noindent\it JCO}\\
The evolution tracks (top left panel) and age-normalised timescale curves (top right panel) of the surface species JCO are shown in Fig.~\ref{JX_IM_surf}. Unlike its gas-phase counterpart, the grain motion effect appears from the very beginning of the modeling. The age-normalised timescale plot shows that its abundance is solely dictated by the dominant grain accretion process at early ages (see the solid curves of ice growth timescale $\tilde{t}_{\rm gro}$ in this panel), which naturally explains the early appearance of the grain motion effect. At ages later than $10^6$\,yr, both the grain accretion and desorption processes become dominant and the desorption process (see the ice evaporation timescale curve $\tilde{t}_{\rm eva}$ in the figure) slightly overtakes the accretion process ($\tilde{t}_{\rm gro}$ curve), so that the overabundance of JCO ice begins to diminish. Some other typical surface species like JC, JN, JO and JCN show similar grain motion effects through the same mechanism.

{\noindent\it JNO}\\
The evolution tracks (left panel of second row) and age-normalised timescale curves (right panel of second row) of the surface species JNO are shown in Fig.~\ref{JX_IM_surf}. JNO illustrates another typical case of the grain motion effect. The JNO abundance is enhanced from the very beginning and an existing sudden abundance jump down around $10^5-10^6$\,yr is also delayed by the grain motion effect. The age-normalised timescale plot  shows that its abundance evolution is always dominated by surface chemical reactions. RRT plot analysis (not shown here) tells us that the dominant production reaction in the early stage is the one between JN and JO whose abundances are directly enhanced by the inclusion of grain motion, which explains the enhancement of JNO abundance.

The sudden abundance drop of JNO is the consequence of decreasing supply of accreted N and O atoms (consumed by accretion onto grains and reaction with other neutrals), the quick consumption of its building blocks JN and JO by other surface reactions, and the consumption of JNO itself in making more complex surface species (dominantly reacts with JH to form JHNO). The RRT plot analysis also indicates that the JNO abundance jump time is a dividing point of two JNO production modes before which it is dominated by the nearly one directional reaction JN + JO $\rightarrow$ JNO, while after which it shifts to the bi-directional equilibrium reaction JNO + JH $\leftrightarrow$ JHNO because the limited surface JN and JO have been used up. Thus, when the grain accretion is enhanced by the grain motion, more O and N atoms are transferred from gas-phase reaction channels to grain accretion channel. Consequently, more surface JN and JO atoms make the first JNO production mode holds for a longer time before the sudden jump. The grain motion effect disappears when the surface chemical network reaches its equilibrium state after about $10^8$\,yr. Some other species like JCN and JO2 also belong to this case.

{\noindent\it JH}\\
The evolution tracks (left panel of third row) and age-normalised timescale curves (right panel of third row) of the surface atomic hydrogen JH are shown in Fig.~\ref{JX_IM_surf}. JH illustrates a unique case where the enhancement of accretion due to grain motion effect significantly reduces its surface abundance immediately after the start of the modeling. Its abundance is jointly controlled by the balance between the equally fast accretion and chemical consumption processes, which can be seen from the overlap of their age-normalised timescale curves (the $\tilde{t}_{\rm gro}$ curves in solid and $\tilde{t}_{\rm con}$ curves in dotted lines) in the right panel of the third row of the figure (note that the two pairs of solid and dotted curves are overlapped so nicely that they can hardly be differentiated in the figure). Thus, the decrease of the JH abundance after the inclusion of grain motion is due to the fact that the grain motion does not increase the accretion of H atom because of its light mass but enhances the abundances of many heavier active surface destroyers of JH.

 An existing sudden jump up of the JH abundance around $10^5-10^6$\,yr is also delayed by the grain motion effect. The jump marks the change of its consumption mode from reacting with simple surface species directly accreted from gas (dominantly JO) to desorption and formation of JH$_2$. Therefore, the faster accretion of species from gas (such as JO) by the grain motion suppresses the JH abundance for a longer time and delays its jump up.

{\noindent\it JH$_2$O}\\
The evolution tracks (bottom left panel) and age-normalised timescale curves (bottom right panel) of the surface water ice JH$_2$O are shown in Fig.~\ref{JX_IM_surf}. The water ice shows little response to the inclusion of grain motion. This is because it is mainly formed on grain surface by a chain of hydrogen addition reactions. One half of the material used to make water ice (JO atom) is surely enhanced by the grain motion effect, while the other half (JH atom) is reduced due to the same effect, as showed above. Therefore, the two opposite effects cancel out and leave the surface water ice abundance insensitive to the inclusion of grain motion. Similar mechanism is also applicable to ammonia, H$_2$S and SiH$_4$ ice.

\subsubsection{Ionic species}
\label{chem_example_ion}

For the ionic species, we have introduced detailed charge accretion process with grains in Sect.~\ref{charging}, which is the only channel to interact with grains. The inclusion of grain motion generally tend to reduce the neutralisation rate in the DC(v) models. As will be shown below, however, the neutralisation process turns out to be always unimportant in the modeling of this work, mainly because the adopted grain radius of $0.1\mu$m is so large that the number density of grains is too low. In future work when we will consider a full distribution of grain sizes (with a larger number of smaller grains), the grain charge neutralisation process may become important to the ion abundance evolution. Therefore, the evolution of the ionic species abundance is mainly controlled by their ionisation and electron recombination processes in the gas phase. To understand how the grain motion indirectly alter the ion abundance evolution, we need to invoke the reaction rate tracing (RRT) diagram described in Sect.~\ref{RRT}.

\begin{figure*}
\centering
\includegraphics[angle=0,scale=.91]{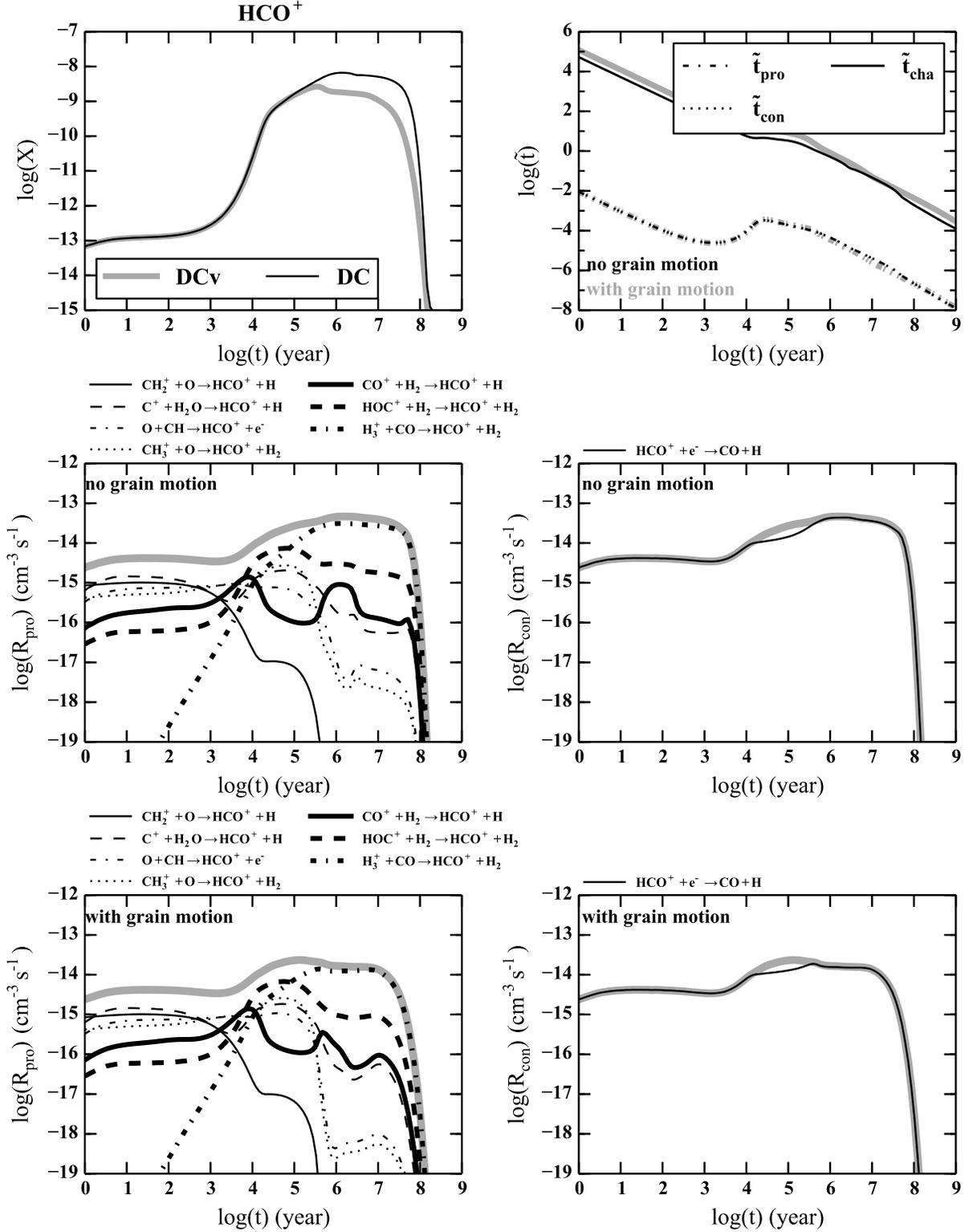}
\caption{Reaction rate tracing (RRT) diagram of HCO$^+$ in the DC(v) cloud models (see the definition of the RRT diagram in Sect.~\ref{RRT}). This diagram is the combination of the abundance evolution plot and age-normalised timescale plot in Fig.~\ref{X_IM_neu} (the first row) with additional reaction rate tracing plots (the middle and bottom rows for the DC and DCv models respectively). The age-normalised timescales (chemical production and consumption timescales, and charge accretion timescale) in the top right panel are defined in Eqs.~(\ref{tnormpro}, \ref{tnormcon}, \ref{tnormneu_ion}). In the middle row, the two panels are the RRT plots of leading production (left) and consumption (right) processes. Here the total rates are plotted in a thick gray solid line, while the individual rates of the major contributing processes are shown with lines of various types and widths, indicating the main drivers of the chemical evolution. The corresponding chemical formula of each plotted reaction is given on the top of each RRT plot. The importance criteria used to select the leading processes is set to $\alpha=0$ (see its definition in Sect.~\ref{RRT}) in this work. This figure is discussed in Sect.~\ref{chem_example_ion}. }
\label{RRT_HCO}
\end{figure*}

{\noindent\it HCO$^+$}\\
The RRT diagram of HCO$^+$ is shown in Fig.~\ref{RRT_HCO}. The upper left panel shows that the grain motion effect begins to show up as the reduction of the HCO$^+$ abundance after an age of $10^5$\,yr. The age-normalised timescale curves (top right panel) tells us that the chemical production and consumption processes are always chemically important and close to balanced with each other, while the ion-grain neutralisation is always unimportant (with much longer timescales $\tilde{t}_{\rm cha}$; see the solid lines in the figure). In the RRT plots (the second and bottom rows), the chemical production of HCO$^+$ is driven by a collection of ion-neutral reactions in the early stages before $\sim 10^4$\,yr and solely by the reaction between H$_3^+$ and CO (the thick black dash-dotted curves in middle left and bottom left panels of the figure) after then, while the HCO$^+$ molecules are destroyed mainly by the recombination processes (the black solid curves in the middle right and bottom right panels) at all time.

The lower HCO$^+$ abundance found in the DCv model (with grain motion effect) in the later stages is mainly the consequence of the significant reduction in abundance of one of its building blocks, gas-phase CO, by the grain motion effect. As we will discuss below, the other building block, H$_3^+$, actually has enhanced abundances due to the grain motion effect during the same evolutionary stage which partially cancels out the opposite effect in CO. However, the decrease in CO abundance is larger and the net effect is the decrease of the HCO$^+$ abundance, as seen from the figure.

\begin{figure*}
\centering
\includegraphics[angle=0,scale=.93]{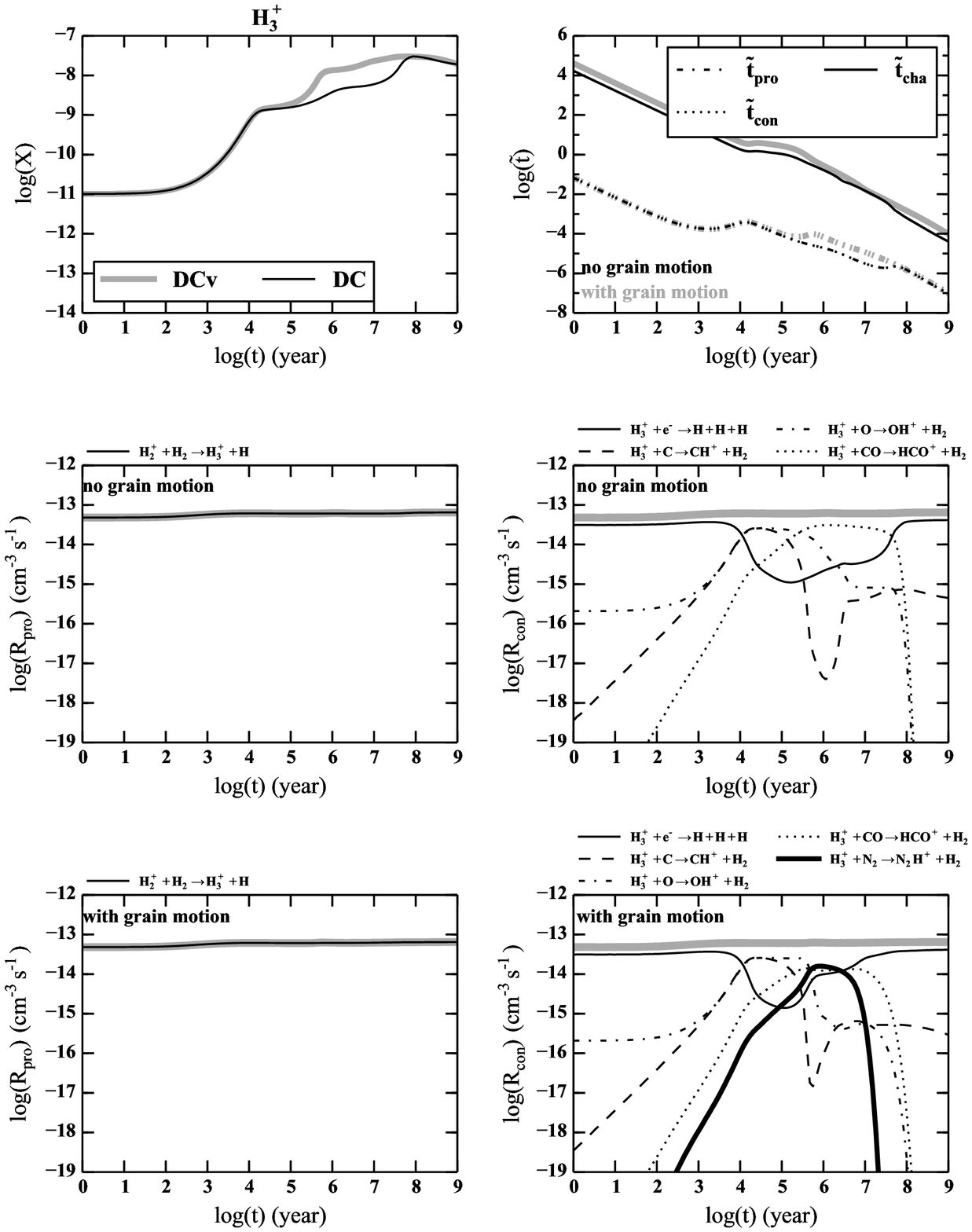}
\caption{Reaction rate tracing (RRT) diagram of H$_3^+$ (similar as Fig.~\ref{RRT_HCO}). This figure is discussed in Sect.~\ref{chem_example_ion}. }
\label{RRT_H3+}
\end{figure*}

{\noindent\it H$_3^+$}\\
This ion represents an opposite case to HCO$^+$, which reveals an interesting general recipe in understanding the chemical effects of a key species (see below). Contrary to the HCO$^+$ case, H$_3^+$ shows enhancement in its abundance in the late evolutionary stages due to grain motion effect, as can be seen in the top left panel of its RRT diagram in Fig.~\ref{RRT_H3+}.

The age-normalised timescale plot (top right panel) demonstrates that the chemical production and consumption processes are always important and the two processes are always close to balance with each other. The ion-neutralisation process is again unimportant at all ages.

In the middle and bottom rows, while the production processes are always dominated by the reaction between H$_2^+$ and H$_2$ all the time, the destruction channel of this ion is jointly controlled by the recombination and a collection of ion-neutral reactions. The enhancement of the H$_3^+$ abundance in the later stages is mainly the consequence of the greater loss of some neutral destroyers like CO that are sensitive to the grain motion effect. The decrease of neutral destroyer abundances and the increase of H$_3^+$ abundance balance with each other so that the total destruction rate of H$_3^+$ (middle right and bottom right panels) is almost not altered by the grain motion effect. The invariance of the total destruction rate is also the natural consequence of the stable balance between the total production and destruction rates of H$_3^+$ (comparing the left and right panels in the middle and bottom rows of Fig.~\ref{RRT_H3+}) and the fact that the single dominating production process is nearly not affected by the grain motion effect (see the only listed chemical process H$_2^+$+H$_2\rightarrow $ H$_3^+$+H on top of the middle and bottom left panels of the figure).

The HCO$^+$ and H$_3^+$ form an pair of opposite cases. The former has a single dominating consumption process HCO$^+$+e$^-$ $\rightarrow$ CO+H and more than one similarly important production reactions, while the latter has a single dominating production process and more than one similarly important consumption reactions; the former shows a decrease of abundance, while the later shows an increase of abundance; the former has both total production and total consumption rates decreased by the enhanced loss of gas-phase CO due to grain motion effect, while the latter shows almost no change in either total production or total consumption rates after the inclusion of grain motion.

The opposite behaviors of HCO$^+$ and H$_3^+$ stem from the fact that the major abundance change induced by grain motion occurs to one of the building blocks of HCO$^+$, while it occurs to some destroyers of H$_3^+$.
As a rule of thumb, if the change of abundance occurs to a major building block of a
considered species, both the total abundance, production rate and the abundance of the
considered species will vary proportionally; if the change occurs to one of the major destroyers
of a species, the abundance of the considered species will vary inversely,
while the total destruction rate may keep almost unchanged because the major destroyer
and the considered species itself vary in an opposite sense and cancel each other out.
\begin{figure*}
\centering
\includegraphics[angle=0,scale=.93]{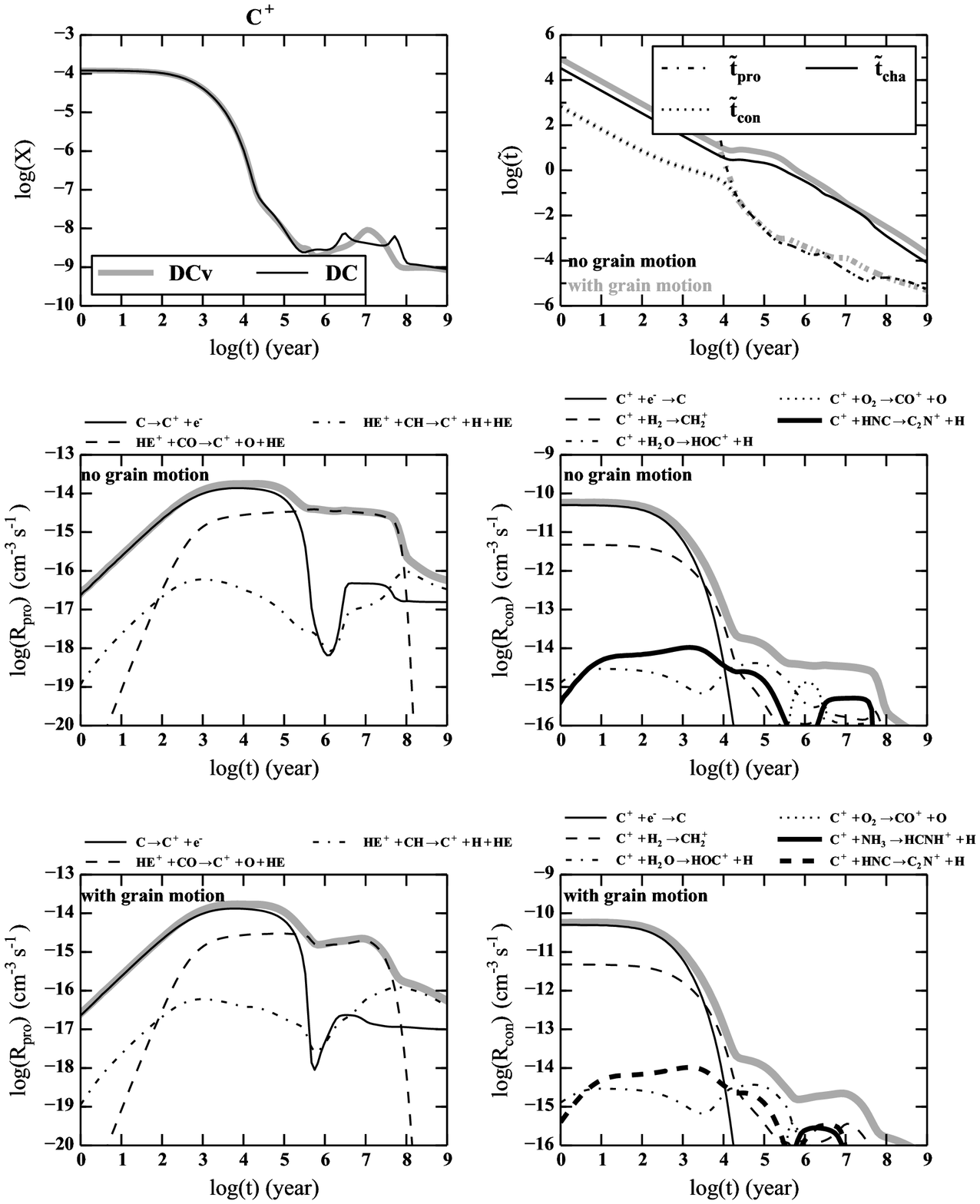}
\caption{Reaction rate tracing (RRT) diagram of C$^+$ (similar as Fig.~\ref{RRT_HCO}). See the discussions in Sect.~\ref{chem_example_ion}. }
\label{RRT_C+}
\end{figure*}

{\noindent\it C$^+$}\\
This ion represents an intermediate case between that of HCO$^+$ and H$_3^+$. The C$^+$ abundance evolution curves (top left panel of Fig.~\ref{RRT_C+}) show little response to the inclusion of grain motion. Both the production and consumption processes involve important contributions from ion-neutral reactions, while such involvement occurs only to the production processes of HCO$^+$ and only to the consumption processes of H$_3^+$. As can be seen from the second and bottom rows in the RRT diagram of C$^+$ in Fig.~\ref{RRT_C+}, the main contributors beside the ionisation processes to the production rate involves some neutral species such as CO and CH, while the main contributors beside the recombination processes to the consumption rate similarly involve a number of neutral species like H$_2$, H$_2$O, HNC O$_2$ and NH$_3$. Thus the decrease of the neutral abundances due to grain motion has similar effects to both total production and consumption rates of C$^+$ and they cancel out, which explains the insensitivity of the C$^+$ abundance to the grain motion.

\subsection{Molecular cloud (MC) models}
\label{MC0}

\begin{figure}
\centering
\includegraphics[angle=0,scale=.45]{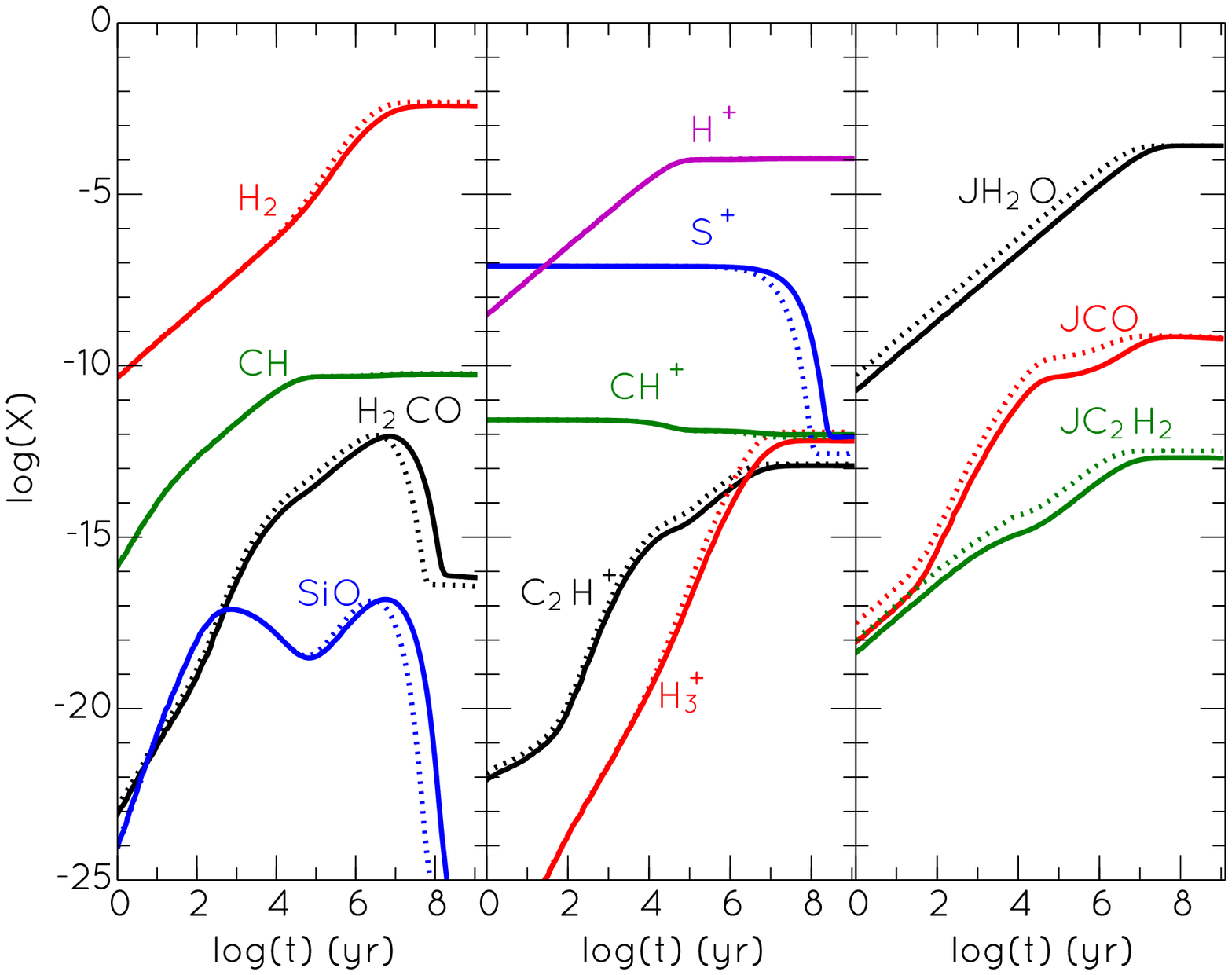}
\caption{Representative cases of grain motion effect to abundance evolution tracks in the MC model for gas-phase neutral species ({\em left}), ionic species ({\em middle}) and surface species ({\em right}). Solid and dotted curves are for models without and with grain motion effect, respectively. See the discussions in Sect.~\ref{MC0}. {\em (A color version of this figure is available online only.)}}
\label{fig7}
\end{figure}

Compared with the DC(v) models, the MC(v) models have much lower gas and grain number densities, a lower extinction, and slightly higher gas and grain temperatures. As a result, the MC(v) models have significantly simpler chemical composition, with fewer chemical species whose maximum density is higher than $10^{-15}$\,cm$^{-3}$. There is almost no abundant species more complex than four atoms (with the only exception of C$_3$H$_2$).

The grain motion effects in the chemical abundances are generally smaller in the MC(v) models than in the DC(v) models. Several representative cases of the grain motion effects for selected neutral, ionic and surface species are shown in the three panels of Fig.~\ref{fig7}. Different from the general decreasing trends of both neutral and ionic abundances in the DC(v) models, many neutral and ionic species in the MC(v) models show abundance enhancement by the grain motion effect (but only by smaller factors of up to only two, compared with the much larger factors of up to 2-3 orders of magnitude in the DC(v) models). Particularly, some gas-phase neutral species (e.g, SiC, HCN, CH$_4$, C$_3$H, H$_2$CO and CS) and ionic species (e.g., C$_2$S$^+$, H$_2$CO$^+$, C$_3$$^+$, C$_2$N$^+$, CH$_3$$^+$, and C$_2$H$^+$) start to show the grain motion effect from the very beginning of the modeling ($t=0$), which do not happen in the DC(v) models. Similar as in DC(v) models, the sudden jumps in the chemical evolution curves also shift to slightly earlier ages (by smaller factors of up to only three, compared to the larger time shifts of up to 10 times in the DC(v) models). Only few abundant ionic species such as P$^+$, H$^+$, He$^+$ in the MC(v) models are insensitive to the grain motion effect. All of the surface species show prominent enhancement of abundance (by factors up to 10) from the beginning of the modeling after the inclusion of the grain motion (see the examples of JH$_2$O, JCO, JC$_2$H$_2$ in the right panel of the figure).

\subsection{Cold Neutral Medium (CNM) models}
\label{CNM0}

\begin{figure}
\centering
\includegraphics[angle=0,scale=.45]{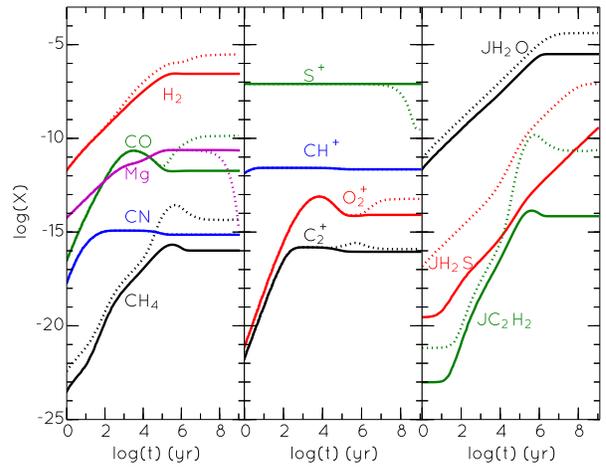}
\caption{Similar as Fig.~\ref{fig7} but for the CNM model. See the discussions in Sect.~\ref{CNM0}. {\em (A color version of this figure is available online only.)}}
\label{fig8}
\end{figure}
With even lower gas and grain number densities, full exposure to the interstellar radiation field, and higher gas and dust temperatures in the CNM(v) models, the number of abundant gas-phase species (with density higher than $10^{-15}$\,cm$^{-3}$) is even smaller, which is true for both neutral and ionic gaseous species. Almost all of the neutral and ionic gas-phase species are composed of no more than two atoms, with only few exceptions (CH$_3$, CH$_4$, H$_2$O and O$_3$).

To help understand the grain motion effects to the chemical abundances, several representative cases for selected neutral, ionic and surface species are shown in Fig.~\ref{fig8}. Similar as in the MC(v) models, most of the gas-phase species (both neutral and ionic) in the CNM(v) models have their abundances enhanced by the grain motion effect, although a larger number of both neutral and ionic species in the CNM(v) models are insensitive to the grain motion effects. The gas-phase abundance enhancement factors in CNM(v) models (10-100) are significantly larger than that in MC(v) models but comparable to the abundance reduction factors in the DC(v) models. Several neutral species show the grain motion effects from the very beginning of the modeling, while several others show this effect from a later model age of $10^2-10^4$\,yr (see the examples of CH$_4$, H$_2$ and CO in the left panel of the figure). Differently, ionic species start to show the grain motion effect only from later ages of $10^4-10^6$\,yr (see the examples of C$_2^+$, O$_2^+$ and S$^+$ in the middle panel of the figure). Particularly, the surface species in the CNM(v) models show the largest abundance enhancement that amounts to 2-6 or even more orders of magnitude (see the examples of JH$_2$O, JH$_2$S, JC$_2$H$_2$ in the right panel of the figure).

\subsection{Discussion on the major differences among the three cloud models}
\label{remarks}

\begin{figure*}
\centering
\includegraphics[angle=0,scale=0.7]{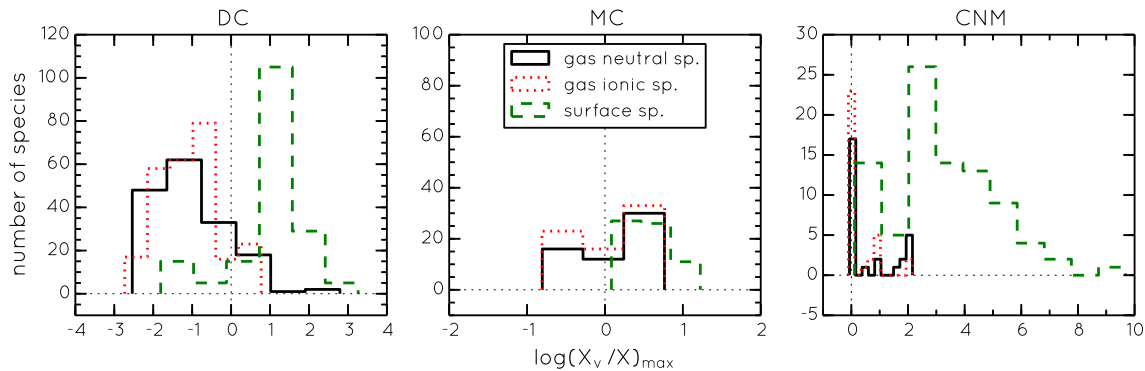}
\caption{Histogram of the maximum abundance change $\log (X_{\rm v}/X)_{\rm max}$ of species due to grain motion effect. The time periods when the chemical evolution curves show a sharp abundance jump has been excluded from the computation of the maximum abundance change. The DC ({\em left}), MC ({\em middle}) and CNM ({\em right}) models are shown in separate panels. The black solid, red dotted and green dashed lines are for neutral, ionic and surface species, respectively. See the discussions in Sect.~\ref{remarks}. {\em (A color version of this figure is available online only.)}}
\label{fighist}
\end{figure*}
Since distinctions have been found in the chemical effects of grain motion among the DC, MC and CNM models, we highlight the most salient differences and discuss the possible reasons in this subsection. To ease our discussions below, we compute the histograms of grain motion effects of all abundant species (with a number density higher than $10^{-15}$\,cm$^{-3}$) in Fig.~\ref{fighist}. An abundance change due to grain motion is defined as $\log (X_{\rm v}/X)_{\rm max}$, where $X_{\rm v}$ and $X$ represent the abundances of the species X in the models with and without grain motion. Gas-phase neutral and ionic species and surface species are considered separately to reveal their different responses to the grain motion. Those time periods with sharp abundance jumps in the chemical evolution curves are excluded from the statistics. We find respectively 593, 194 and 149 chemical species in the DC(v), MC(v)  and CNM(v) models that are abundant enough to enter our statistics, which confirms our earlier conclusions that the chemical composition is the most complex in the DC model and the simplest in the CNM model. This has been explained earlier mainly by the lower density (more difficult to form complex species) and higher ISRF (easier to destroy the complex species) in the MC and CNM models.

First we compare the neutral (black solid lines) and ionic (red dotted lines) species in the three panels of Fig.~\ref{fighist} and find that most of the gas-phase species show abundance decreases in the DC(v) models (with $\log (X_{\rm v}/X)_{\rm max}<0$), slightly more abundance increases than decreases in the MC(v) models and only non-negative abundance changes in the CNM(v) models. This agrees with our earlier findings that the grain motion effect to the gas-phase species is generally to bring down the abundances of both the neutral and ionic species in the DC(v) models, but inversely push them up in the MC(v) and CNM(v) models (although exceptional cases coexist). This difference stems from the fact that the gas-phase abundances are dominated by gas-phase chemical reactions in the MC(v) models but are strongly affected or even overwhelmed by the formation of neutral species on the grains plus ensuing desorption and gas-phase processes in the MC(v)and CNM(v) models. For example, in the CNM(v) models, N$_2$, O$_3$ and CH$_4$ are dominantly formed on grains so that all of them and their related species such as CH$_3$ and N$_2^+$ start to show abundance enhancement by grain motion since $t=0$. Some other species such as H$_2$, O$_2$, CH and CO are still formed mainly through gaseous chemical reactions in the CNM(v) models, but the surface formation channel still becomes relative important or even dominant at least in some period of the evolution, so that all of them and their related species such as H$_2^+$, HCL$^+$, NH$^+$, and C$_2^+$ show abundance increase (as the response to the grain motion effects) during these periods.

The figure also reveals that the extent of the grain motion effect to the gas-phase abundances is the smallest in MC(v) models (by factors less than 10) and larger in DC(v) and CNM(v) models (by factors up to 2-3 orders of magnitude). This phenomenon can be jointly explained by three facts: 1) at low temperature and high gas density, the dominant grain motion effect is to reduce the gas-phase abundances, which is most prominent in the DC(v) models, as can be seen from the left panel of Fig.~\ref{fig_kacc}; 2) at higher temperature and low gas density, the dominant grain motion effect is to enhance the gas-phase abundances via grain surface formation channel which is the most efficient in the CNM(v) models because of the highest grain motion speed in this case, as can be seen in Table~\ref{tab_phy_models} and the right panel of Fig.~\ref{fig_kacc}; 3) the opposite trends of grain motion effect to reduce or enhance the gas-phase abundances are not only weaker in the MC(v) models but cancel each other to some degree. Therefore, the MC(v) models serve as good intermediate cases.

Comparing the surface species (green dashed lines) among the three panels in Fig.~\ref{fighist}, one can easily see that the inclusion of grain motion generally tends to enhance the surface abundances (with $\log (X_{\rm v}/X)_{\rm max}>0$) in all the cloud models (with only few exceptions in the DC(v) models in the left panel). However, the enhancement is the largest in the CNM(v) models (by factors of mostly $10^2-10^6$ and even up to $10^9$), intermediate in the DC(v) models (by factors up to $10^3$), and smallest in the MC(v) models  (by factors only up to $10$). This is the consequence of the variation of physical conditions among the three cloud models. The amount of neutral gas decreases from the DC(v) to MC(v) to CNM(v) models while the amount of ionised gas increases along the same sequence due to the different gas densities and dust extinctions. The immediate grain motion effect is to enhance the accretion of neutral gas onto all grains (dominant in the MC model) and to increase the accretion of cations onto positively charged grains (dominant in the CNM model). On the one hand, the enhancement by grain motion to the neutral accretion rates is not only weaker but less important in the MC(v) models than in the DC(v) models, because of the lower amount of neutral gas in the former models and their lower sensitivity to grain motion due to higher gas temperature, as discussed in Sect.~\ref{Racc_example_neu} and the left panel of Fig.~\ref{fig_kacc}. On the other hand, the enhancement by grain motion to the cation accretion rates is also weaker and less important in the MC(v) models than in the CNM(v) models, because of the lower amount of ionised gas in the former models and their smaller change in accretion rate coefficients due to slower grain motions, as discussed in Sect.~\ref{Racc_example_ion} and the right panel of Fig.~\ref{fig_kacc}.

\section{Summary}
\label{summary}

In turbulent interstellar clouds, dust grains are usually moving through the gaseous medium, which can have a consequence to the cloud chemistry by enhancing the grain accretion rates of neutral species, increasing the accretion of ions onto grains of the same type of charge, and reducing the charge neutralisation rates between ions and oppositely charged grains. The chemical effects of grain motions are tested using our new gas-grain chemical code `ggchem'. In this pioneering work, we only consider a single grain radius of $1.0\times 10^{-5}$\,cm and adopt the corresponding average grain motion speed from the work of \citet{yan2004}. Analysis of relevant timescales is performed and special tool plots such as `age-normalised timescale plots' and `reaction rate tracing (RRT) diagrams' are used to help understand the numerical results. The grain motion is found to have important effects to the cloud chemistry.

In the typical conditions of dark clouds (DC model), the following interesting facts on how and when the grain motions begin to take effect in the cloud chemistry are found:
\begin{enumerate}{
\item{The grain motion effects to the chemical abundance evolution can be classified into two categories: in those time periods when the abundance evolves slowly, the grain motion effects show up as change of the abundances to higher or lower values; in those moments when the abundances are experiencing sharp jumps, the grain motion effects appear as the shift of the jumps to earlier or later ages.}
\item{Both the neutral and ionic gas-phase abundances generally tend to be reduced by the grain motion effects by factors up to 2-3 orders of magnitude at later stages of the chemical evolution, while the surface icy species abundances generally tend to be enhanced by factors up to 1-2 orders of magnitude from the beginning of the modeling, although a few exceptional cases exist.}
\item{Most gas-phase species that have sharp abundance jumps in their chemical evolution curves show the jumps at about ten times earlier times, while most surface species that have similar sharp abundance jumps delay the jumps to later times by factors of several.}
\item{There exist some exceptional cases in which some species are insensitive to the inclusion of grain motions and some other species show opposite behaviors than the majority. For example, some gas-phase species can show enhanced abundance (e.g., O$_3$ and H$_3^+$) and some surface species can have reduced abundances (e.g., JH) due to the grain motion effect.}
}
\end{enumerate}
The mechanisms behind these behaviors have been explained in detail by applying the aforementioned tool plots to exemplar species such as CO, C$_2$, SO$_3$, JCO, JNO, JH, JH$_2$O, HCO$^+$, H$_3^+$ and C$^+$ in the DC(v) models.

The grain motion effects in the CNM(v) models represent the opposite extreme cases to the DC(v) models. In this case, while many gas-phase species are insensitive to the addition of grain motion, almost all other gas-phase neutrals, ions and surface icy species have their abundances greatly enhanced by the grain motion (by factors up to $10^2$ to $10^6$ or even larger). The prominent effect is a consequence of the fact that the additional energy from the grain motion effectively helps the cations to overcome the Coulomb barrier on the positively charged grains and make the accretion of cations onto grains efficient.

The grain motion effect in the MC(v) models behave as intermediate cases between the DC(v) and CNM(v) models. Beside some insensitive species, the abundances of most other neutral, ionic and surface species are only slightly enhanced or reduced by the inclusion of grain motion. The extent of abundance change is smaller than both DC(v) and CNM(v) models.

This work has demonstrated that the turbulent grain motion can significantly alter the interstellar chemistry in the typical interstellar conditions.

\section*{Acknowledgements}

JH owes great debt to Dr. T.I. Hasegawa for his help to many aspects of this research. We also acknowledge Dr. Demitry A. Semenov for the helpful correspondences upon code benchmarking and grain-ion interaction and Prof. Aigen Li for offering us his Fortran code and dust optical data for computing grain absorption coefficients. JH thanks the support of the NSFC grant No. 11173056. HY acknowledges the support by the NSFC grant 11473006.

\bibliography{dust-velocity}

\begin{thebibliography}{}

\bibitem[\protect\citeauthoryear{{Acharyya}, {Hassel} \& {Herbst}}{{Acharyya}
  et~al.}{2011}]{acha2011}
{Acharyya} K.,  {Hassel} G.~E.,    {Herbst} E.,  2011, \apj, 732, 73

\bibitem[\protect\citeauthoryear{{Arons} \& {Max}}{{Arons} \&
  {Max}}{1975}]{aron1975}
{Arons} J.,  {Max} C.~E.,  1975, \apjl, 196, L77

\bibitem[\protect\citeauthoryear{{Barzel} \& {Biham}}{{Barzel} \&
  {Biham}}{2007}]{barz2007}
{Barzel} B.,  {Biham} O.,  2007, \apjl, 658, L37

\bibitem[\protect\citeauthoryear{{Biham}, {Furman}, {Pirronello} \&
  {Vidali}}{{Biham} et~al.}{2001}]{biha2001}
{Biham} O.,  {Furman} I.,  {Pirronello} V.,    {Vidali} G.,  2001, \apj, 553,
  595

\bibitem[\protect\citeauthoryear{{Boulanger}, {Abergel}, {Bernard}, {Burton},
  {Desert}, {Hartmann}, {Lagache} \& {Puget}}{{Boulanger}
  et~al.}{1996}]{boul1996}
{Boulanger} F.,  {Abergel} A.,  {Bernard} J.-P.,  {Burton} W.~B.,  {Desert}
  F.-X.,  {Hartmann} D.,  {Lagache} G.,    {Puget} J.-L.,  1996, \aap, 312, 256

\bibitem[\protect\citeauthoryear{{Caselli}, {Hasegawa} \& {Herbst}}{{Caselli}
  et~al.}{1998}]{case1998}
{Caselli} P.,  {Hasegawa} T.~I.,    {Herbst} E.,  1998, \apj, 495, 309

\bibitem[\protect\citeauthoryear{{Chang}, {Cuppen} \& {Herbst}}{{Chang}
  et~al.}{2005}]{chan2005}
{Chang} Q.,  {Cuppen} H.~M.,    {Herbst} E.,  2005, \aap, 434, 599

\bibitem[\protect\citeauthoryear{{Chang}, {Cuppen} \& {Herbst}}{{Chang}
  et~al.}{2006}]{chan2006}
{Chang} Q.,  {Cuppen} H.~M.,    {Herbst} E.,  2006, \aap, 458, 497

\bibitem[\protect\citeauthoryear{{Chang}, {Cuppen} \& {Herbst}}{{Chang}
  et~al.}{2007}]{chan2007}
{Chang} Q.,  {Cuppen} H.~M.,    {Herbst} E.,  2007, \aap, 469, 973

\bibitem[\protect\citeauthoryear{{Chang} \& {Herbst}}{{Chang} \&
  {Herbst}}{2012}]{chan2012}
{Chang} Q.,  {Herbst} E.,  2012, \apj, 759, 147

\bibitem[\protect\citeauthoryear{{Charnley}}{{Charnley}}{1998}]{char1998}
{Charnley} S.~B.,  1998, \apjl, 509, L121

\bibitem[\protect\citeauthoryear{{Charnley}, {Tielens} \& {Rodgers}}{{Charnley}
  et~al.}{1997}]{char1997}
{Charnley} S.~B.,  {Tielens} A.~G.~G.~M.,    {Rodgers} S.~D.,  1997, \apjl,
  482, L203

\bibitem[\protect\citeauthoryear{{Dalgarno}}{{Dalgarno}}{2006}]{dalg2006}
{Dalgarno} A.,  2006, Proceedings of the National Academy of Science, 103,
  12269

\bibitem[\protect\citeauthoryear{{Draine}}{{Draine}}{1978}]{drai1978}
{Draine} B.~T.,  1978, \apjs, 36, 595

\bibitem[\protect\citeauthoryear{{Draine} \& {Sutin}}{{Draine} \&
  {Sutin}}{1987}]{drai1987}
{Draine} B.~T.,  {Sutin} B.,  1987, \apj, 320, 803

\bibitem[\protect\citeauthoryear{{Garrod}}{{Garrod}}{2013a}]{garr2013a}
{Garrod} R.~T.,  2013a, \apj, 765, 60

\bibitem[\protect\citeauthoryear{{Garrod}}{{Garrod}}{2013b}]{garr2013b}
{Garrod} R.~T.,  2013b, \apj, 778, 158

\bibitem[\protect\citeauthoryear{{Garrod} \& {Herbst}}{{Garrod} \&
  {Herbst}}{2006}]{garr2006}
{Garrod} R.~T.,  {Herbst} E.,  2006, \aap, 457, 927

\bibitem[\protect\citeauthoryear{{Garrod}, {Vasyunin}, {Semenov}, {Wiebe} \&
  {Henning}}{{Garrod} et~al.}{2009}]{garr2009}
{Garrod} R.~T.,  {Vasyunin} A.~I.,  {Semenov} D.~A.,  {Wiebe} D.~S.,
  {Henning} T.,  2009, \apjl, 700, L43

\bibitem[\protect\citeauthoryear{{Garrod}, {Weaver} \& {Herbst}}{{Garrod}
  et~al.}{2008}]{garr2008}
{Garrod} R.~T.,  {Weaver} S.~L.~W.,    {Herbst} E.,  2008, \apj, 682, 283

\bibitem[\protect\citeauthoryear{{Goldreich} \& {Sridhar}}{{Goldreich} \&
  {Sridhar}}{1995}]{gold1995}
{Goldreich} P.,  {Sridhar} S.,  1995, \apj, 438, 763

\bibitem[\protect\citeauthoryear{{Gould} \& {Salpeter}}{{Gould} \&
  {Salpeter}}{1963}]{goul1963}
{Gould} R.~J.,  {Salpeter} E.~E.,  1963, \apj, 138, 393

\bibitem[\protect\citeauthoryear{{G{\"u}sten}, {Walmsley} \&
  {Pauls}}{{G{\"u}sten} et~al.}{1981}]{gues1981}
{G{\"u}sten} R.,  {Walmsley} C.~M.,    {Pauls} T.,  1981, \aap, 103, 197

\bibitem[\protect\citeauthoryear{{Hasegawa} \& {Herbst}}{{Hasegawa} \&
  {Herbst}}{1993a}]{hase1993a}
{Hasegawa} T.~I.,  {Herbst} E.,  1993a, \mnras, 261, 83

\bibitem[\protect\citeauthoryear{{Hasegawa} \& {Herbst}}{{Hasegawa} \&
  {Herbst}}{1993b}]{hase1993b}
{Hasegawa} T.~I.,  {Herbst} E.,  1993b, \mnras, 263, 589

\bibitem[\protect\citeauthoryear{{Hasegawa}, {Herbst} \& {Leung}}{{Hasegawa}
  et~al.}{1992}]{hase1992}
{Hasegawa} T.~I.,  {Herbst} E.,    {Leung} C.~M.,  1992, \apjs, 82, 167

\bibitem[\protect\citeauthoryear{{Hincelin}, {Chang} \& {Herbst}}{{Hincelin}
  et~al.}{2015}]{hinc2015}
{Hincelin} U.,  {Chang} Q.,    {Herbst} E.,  2015, \aap, 574, A24

\bibitem[\protect\citeauthoryear{{Hirashita} \& {Yan}}{{Hirashita} \&
  {Yan}}{2009}]{hira2009}
{Hirashita} H.,  {Yan} H.,  2009, \mnras, 394, 1061

\bibitem[\protect\citeauthoryear{{Hoang} \& {Lazarian}}{{Hoang} \&
  {Lazarian}}{2012}]{Hoang2012}
{Hoang} T.,  {Lazarian} A.,  2012, \apj, 761, 96

\bibitem[\protect\citeauthoryear{{Hollenbach} \& {Salpeter}}{{Hollenbach} \&
  {Salpeter}}{1971}]{holl1971}
{Hollenbach} D.,  {Salpeter} E.~E.,  1971, \apj, 163, 155

\bibitem[\protect\citeauthoryear{{Indriolo}, {Geballe}, {Oka} \&
  {McCall}}{{Indriolo} et~al.}{2007}]{Ind2007}
{Indriolo} N.,  {Geballe} T.~R.,  {Oka} T.,    {McCall} B.~J.,  2007, \apj,
  671, 1736

\bibitem[\protect\citeauthoryear{{Indriolo} \& {McCall}}{{Indriolo} \&
  {McCall}}{2012}]{Ind2012}
{Indriolo} N.,  {McCall} B.~J.,  2012, \apj, 745, 91

\bibitem[\protect\citeauthoryear{{Ivlev}, {Lazarian}, {Tsytovich}, {de
  Angelis}, {Hoang} \& {Morfill}}{{Ivlev} et~al.}{2010}]{Ivle2010}
{Ivlev} A.~V.,  {Lazarian} A.,  {Tsytovich} V.~N.,  {de Angelis} U.,  {Hoang}
  T.,    {Morfill} G.~E.,  2010, \apj, 723, 612

\bibitem[\protect\citeauthoryear{{Kalv{\= a}ns} \& {Shmeld}}{{Kalv{\= a}ns} \&
  {Shmeld}}{2010}]{kalv2010}
{Kalv{\= a}ns} J.,  {Shmeld} I.,  2010, \aap, 521, A37

\bibitem[\protect\citeauthoryear{{Kruegel} \& {Walmsley}}{{Kruegel} \&
  {Walmsley}}{1984}]{krue1984}
{Kruegel} E.,  {Walmsley} C.~M.,  1984, \aap, 130, 5

\bibitem[\protect\citeauthoryear{{Lada}, {Thronson} Jr., {Smith}, {Harper},
  {Keene}, {Loewenstein} \& {Smith}}{{Lada} et~al.}{1981}]{lada1981}
{Lada} C.~J.,  {Thronson} Jr. H.~A.,  {Smith} H.~A.,  {Harper} D.~A.,  {Keene}
  J.,  {Loewenstein} R.~F.,    {Smith} J.,  1981, \apjl, 251, L91

\bibitem[\protect\citeauthoryear{{Lazarian} \& {Yan}}{{Lazarian} \&
  {Yan}}{2002}]{Laza2002}
{Lazarian} A.,  {Yan} H.,  2002, \apjl, 566, L105

\bibitem[\protect\citeauthoryear{{Le Bourlot}, {Le Petit}, {Pinto}, {Roueff} \&
  {Roy}}{{Le Bourlot} et~al.}{2012}]{lebo2012}
{Le Bourlot} J.,  {Le Petit} F.,  {Pinto} C.,  {Roueff} E.,    {Roy} F.,  2012,
  \aap, 541, A76

\bibitem[\protect\citeauthoryear{{Lepp}}{{Lepp}}{1992}]{Lepp1992}
{Lepp} S.,  1992, in {Singh} P.~D.,  ed., Astrochemistry of Cosmic Phenomena
  Vol.~150 of IAU Symposium, {The Cosmic-Ray Ionization RATE*}.
p.~471

\bibitem[\protect\citeauthoryear{{Lepp} \& {Dalgarno}}{{Lepp} \&
  {Dalgarno}}{1988}]{lepp1988}
{Lepp} S.,  {Dalgarno} A.,  1988, \apj, 324, 553

\bibitem[\protect\citeauthoryear{{Li} \& {Draine}}{{Li} \&
  {Draine}}{2001}]{Li2001}
{Li} A.,  {Draine} B.~T.,  2001, \apj, 554, 778

\bibitem[\protect\citeauthoryear{{Mathis}, {Mezger} \& {Panagia}}{{Mathis}
  et~al.}{1983}]{math1983}
{Mathis} J.~S.,  {Mezger} P.~G.,    {Panagia} N.,  1983, \aap, 128, 212

\bibitem[\protect\citeauthoryear{{McCall}, {Huneycutt}, {Saykally}, {Geballe},
  {Djuric}, {Dunn}, {Semaniak}, {Novotny}, {Al-Khalili}, {Ehlerding},
  {Hellberg}, {Kalhori}, {Neau}, {Thomas}, {{\"O}sterdahl} \&
  {Larsson}}{{McCall} et~al.}{2003}]{Mcc2003}
{McCall} B.~J.,  {Huneycutt} A.~J.,  {Saykally} R.~J.,  {Geballe} T.~R.,
  {Djuric} N.,  {Dunn} G.~H.,  {Semaniak} J.,  {Novotny} O.,  {Al-Khalili} A.,
  {Ehlerding} A.,  {Hellberg} F.,  {Kalhori} S.,  {Neau} A.,  {Thomas} R.,
  {{\"O}sterdahl} F.,    {Larsson} M.,  2003, \nat, 422, 500

\bibitem[\protect\citeauthoryear{{Omont}}{{Omont}}{1986}]{omon1986}
{Omont} A.,  1986, \aap, 164, 159

\bibitem[\protect\citeauthoryear{{Sargent}, {van Duinen}, {Nordh}, {Fridlund},
  {Aalders} \& {Beintema}}{{Sargent} et~al.}{1983}]{sarg1983}
{Sargent} A.~I.,  {van Duinen} R.~J.,  {Nordh} H.~L.,  {Fridlund} C.~V.~M.,
  {Aalders} J.~W.~G.,    {Beintema} D.,  1983, \aj, 88, 88

\bibitem[\protect\citeauthoryear{{Schutte} \& {Greenberg}}{{Schutte} \&
  {Greenberg}}{1991}]{schu1991}
{Schutte} W.~A.,  {Greenberg} J.~M.,  1991, \aap, 244, 190

\bibitem[\protect\citeauthoryear{{Semenov}, {Hersant}, {Wakelam}, {Dutrey},
  {Chapillon}, {Guilloteau}, {Henning}, {Launhardt}, {Pi{\'e}tu} \&
  {Schreyer}}{{Semenov} et~al.}{2010}]{seme2010}
{Semenov} D.,  {Hersant} F.,  {Wakelam} V.,  {Dutrey} A.,  {Chapillon} E.,
  {Guilloteau} S.,  {Henning} T.,  {Launhardt} R.,  {Pi{\'e}tu} V.,
  {Schreyer} K.,  2010, \aap, 522, A42

\bibitem[\protect\citeauthoryear{{Stantcheva}, {Shematovich} \&
  {Herbst}}{{Stantcheva} et~al.}{2002}]{stan2002}
{Stantcheva} T.,  {Shematovich} V.~I.,    {Herbst} E.,  2002, \aap, 391, 1069

\bibitem[\protect\citeauthoryear{{Tielens} \& {Hagen}}{{Tielens} \&
  {Hagen}}{1982}]{tiel1982}
{Tielens} A.~G.~G.~M.,  {Hagen} W.,  1982, \aap, 114, 245

\bibitem[\protect\citeauthoryear{{Vasyunin} \& {Herbst}}{{Vasyunin} \&
  {Herbst}}{2013}]{vasy2013}
{Vasyunin} A.~I.,  {Herbst} E.,  2013, \apj, 762, 86

\bibitem[\protect\citeauthoryear{{Vasyunin}, {Semenov}, {Wiebe} \&
  {Henning}}{{Vasyunin} et~al.}{2009}]{vasy2009}
{Vasyunin} A.~I.,  {Semenov} D.~A.,  {Wiebe} D.~S.,    {Henning} T.,  2009,
  \apj, 691, 1459

\bibitem[\protect\citeauthoryear{{Vaupr{\'e}}, {Hily-Blant}, {Ceccarelli},
  {Dubus}, {Gabici} \& {Montmerle}}{{Vaupr{\'e}} et~al.}{2014}]{Vaup2014}
{Vaupr{\'e}} S.,  {Hily-Blant} P.,  {Ceccarelli} C.,  {Dubus} G.,  {Gabici} S.,
     {Montmerle} T.,  2014, \aap, 568, A50

\bibitem[\protect\citeauthoryear{{Viti} \& {Williams}}{{Viti} \&
  {Williams}}{1999}]{viti1999}
{Viti} S.,  {Williams} D.~A.,  1999, \mnras, 305, 755

\bibitem[\protect\citeauthoryear{{Wakelam} \& {Herbst}}{{Wakelam} \&
  {Herbst}}{2008}]{wake2008}
{Wakelam} V.,  {Herbst} E.,  2008, \apj, 680, 371

\bibitem[\protect\citeauthoryear{{Weingartner} \& {Draine}}{{Weingartner} \&
  {Draine}}{2001}]{Wei2001}
{Weingartner} J.~C.,  {Draine} B.~T.,  2001, \apjs, 134, 263

\bibitem[\protect\citeauthoryear{{Yan}}{{Yan}}{2009}]{Yan2009}
{Yan} H.,  2009, \mnras, 397, 1093

\bibitem[\protect\citeauthoryear{{Yan} \& {Lazarian}}{{Yan} \&
  {Lazarian}}{2003}]{YL2003}
{Yan} H.,  {Lazarian} A.,  2003, \apjl, 592, L33

\bibitem[\protect\citeauthoryear{{Yan}, {Lazarian} \& {Draine}}{{Yan}
  et~al.}{2004}]{yan2004}
{Yan} H.,  {Lazarian} A.,    {Draine} B.~T.,  2004, \apj, 616, 895

\end{thebibliography}
\bibliographystyle{mn2e}

\label{lastpage}

\end{document}